\documentclass[lettersize,journal]{IEEEtran}
\usepackage[T1]{fontenc}
\def\endthebibliography{%
  \def\@noitemerr{\@latex@warning{Empty `thebibliography' environment}}%
  \endlist
}
\usepackage{balance}

\usepackage{bm}

\usepackage{cite}

\usepackage[pdftex]{graphicx}

\ifCLASSINFOpdf
\else
\fi
\usepackage{diagbox}
\usepackage{xcolor}
\usepackage{amsmath}

\usepackage[cmintegrals]{newtxmath}
\usepackage{algorithm}

\usepackage{algcompatible}

\usepackage{hyperref}
\usepackage{subfigure}
\usepackage{makecell}
\begin{document}
%
\title{Joint Activity Detection, Channel Estimation, and Data Decoding for Grant-free Massive Random Access}
%
%
%

\author{Xinyu~Bian,~\IEEEmembership{Student Member,~IEEE,}
        Yuyi~Mao,~\IEEEmembership{Member,~IEEE,}
        and~Jun~Zhang,~\IEEEmembership{Fellow,~IEEE}
\thanks{Manuscript received 13 April 2022; revised 28 December 2022; accepted 3 February 2023. This work was supported in part by the General Research Fund (Project No. 15207220) from the Hong Kong Research Grants Council and in part by a start-up fund of the Hong Kong Polytechnic University (Project ID P0038174). This paper was presented in part at the 2021 IEEE International Workshop on Signal Processing Advances in Wireless Communications (SPAWC) \cite{xbian20211}. \emph{(Corresponding author: Yuyi Mao.)}

X. Bian and J. Zhang are with the Department of Electronic and Computer Engineering, the Hong Kong University of Science and Technology, Hong Kong (E-mail: xinyu.bian@connect.ust.hk, eejzhang@ust.hk). Y. Mao is with the Department of Electronic and Information Engineering, the Hong Kong Polytechnic University, Hong Kong (E-mail: yuyi-eie.mao@polyu.edu.hk).}
\thanks{}
\thanks{}}

%
%

\markboth{}%
{Shell \MakeLowercase{\textit{et al.}}: Bare Demo of IEEEtran.cls for IEEE Communications Society Journals}
%



\maketitle

\begin{abstract}
In the massive machine-type communication (mMTC) scenario, a large number of devices with sporadic traffic need to access the network on limited radio resources. Recently, grant-free random access has emerged as a promising mechanism for this challenging scenario, but its potential has not been fully unleashed. In particular, the available auxiliary information has not been fully exploited, including the common sparsity pattern in the received pilot and data signal, as well as the channel decoding information. This paper develops advanced receivers in a holistic manner to improve the massive access performance by jointly designing activity detection, channel estimation, and data decoding. To tackle the algorithmic and computational challenges, a turbo structure is adopted at the joint receiver. For performance enhancement, all the received symbols are utilized to jointly estimate the channel state, user activity, and soft data symbols, which effectively exploits the common sparsity pattern. Meanwhile, the extrinsic information from the channel decoder will assist the joint channel estimation and data detection. To reduce the complexity, a low-cost side information (SI)-aided receiver is also proposed, where the channel decoder provides side information to update the estimates on whether a user is active or not. Simulation results show that the turbo receiver is able to reduce the activity detection, channel estimation, and data decoding errors effectively, supporting twice as many active users compared with a separate design that disregards the common sparsity. In addition, the SI-aided receiver notably outperforms the conventional methods with a relatively low complexity.
\end{abstract}

\begin{IEEEkeywords}
Grant-free massive random access, massive machine-type communication (mMTC), user activity detection, channel estimation, channel coding, approximate message passing (AMP), turbo receiver.
\end{IEEEkeywords}

%
\IEEEpeerreviewmaketitle

\section{Introduction}
%
%
%
%
The upsurge of numerous Internet of Things (IoT) applications, e.g., autonomous vehicles, intelligent robots, smart city, and industry 4.0, is boosting a rapid paradigm shift of wireless communications from connecting people to connecting things \cite{aal2015}. It is estimated by Cisco that the share of machine-to-machine (M2M) communications will increase from 33\% in 2018 to 50\% in 2023 \cite{Cisco}, leading to an unprecedented requirement of ubiquitous and scalable connectivity. In order to fulfill such an urgent need, massive machine-type communications (mMTC) has been identified as one of the key application scenarios of the fifth-generation (5G) cellular networks by the third-generation partnership project (3GPP) \cite{itu}. The most distinctive feature of mMTC is that a huge amount of devices are simultaneously connected to a base station (BS), while only a small proportion of them are active for transmitting a short data packet at each time \cite{cboc2016}. This brings forth significant challenges that cannot be catered with existing wireless technologies. In particular, uplink access in cellular networks is traditionally controlled by grant-based random access (RA) mechanisms, where each user first initiates a RA procedure by transmitting a scheduling request (SR) to its serving BS, and it cannot start data transmission until the request is granted \cite{mha2013}. Nevertheless, due to the limited preamble sequences for SR, grant-based RA suffers from potential collisions of the connection requests when two or more users pick the same sequence, especially with massive concurrent requests \cite{ebj2017}. Although complementary techniques such as power ramping, back-off, and access class barring \cite{njiang2018}, can be utilized to relieve access collision, long access latency and significant signalling overhead will be incurred, which are unfavorable for mMTC \cite{lliu2018,psh2017}. Therefore, more efficient RA mechanisms are needed to meet the stringent latency and reliability requirements of mMTC.

Grant-free RA, which allows users to directly send messages without waiting for access permissions from the BS \cite{xchen2021}, is widely acknowledged as a promising alternative for massive RA. In contrast to grant-based RA where each user selects a random pilot sequence at each time slot, grant-free RA assigns a fixed and unique pilot sequence to each user to enable contention-free uplink access. Nevertheless, the BS does not have knowledge of the set of transmitting (i.e., active) users in grant-free RA, making it arduous to perform accurate channel estimation and data reception. Consequently, detecting the set of transmitting users (i.e., user activity detection) at the BS, becomes a new and critical task for grant-free massive RA \cite{ywu2020}.

\subsection{Related Works and Motivations}
The popular pilot-based grant-free RA protocol is considered in this paper, where the active users transmit their unique pilot sequences using dedicated radio resources followed by data symbols, and the BS identifies the set of active users and decodes their data. Attributed to the massive number of devices and the limited resources for pilot transmission, it is infeasible to assign orthogonal pilot sequences to all the devices, rendering conventional collision avoidance mechanisms not readily applicable. Fortunately, user activity detection in mMTC turns out to be a compressive sensing (CS) problem \cite{donoho2006} thanks to the sporadic data traffic pattern, for which, many efficient algorithms are available \cite{trob1996, jat2004, donoho2010, jtpa2014}. For instance, by formulating user activity detection as a maximum likelihood (ML) estimation problem, low-complexity algorithms based on sample covariance matrices of the received pilot signal were proposed for massive multi-input multi-output (MIMO) systems in \cite{shag2018}. The user activity detection accuracies in massive and cooperative MIMO systems were analyzed in \cite{zchen2019} with the approximate message passing (AMP) algorithm. Nevertheless, early studies on grant-free massive RA only focused on user activity detection but neglected the data reception performance, which motivates the investigations on multi-user detection (MUD) for massive connectivity \cite{bwang2016,cwei2017}. In particular, based on the orthogonal matching pursuit (OMP), a support detection algorithm was proposed for joint activity and data detection in grant-free non-orthogonal multiple access (NOMA) systems in \cite{bwang2016}. A similar problem was tackled by fusing the expectation maximization (EM) algorithm with AMP in \cite{cwei2017}, which leverages prior information of the transmitted data symbols in addition to the sparse user activity pattern. However, these works assume full channel state information (CSI) available at the BS, which is idealized and impractical for grant-free massive RA.

To bridge this gap, joint activity detection and channel estimation (JADCE) becomes an emerging theme for grant-free massive RA. It was shown in \cite{zchen2018} that JADCE can be formulated as a single measurement vector (SMV) and multiple measurement vector (MMV) CS problem with single- and multi-antenna BS, respectively, and both problems can be efficiently solved by AMP. The false alarm and missed detection probabilities were also characterized in \cite{zchen2018}. Interestingly, a subsequent investigation \cite{lliuwyu2018} revealed that the activity detection error can be made arbitrarily small with sufficient BS antennas. Besides, characteristics of the wireless channels have been utilized together with the sparse user activity pattern to promote the accuracy of JADCE \cite{mke2020,ycheng2021}. Moreover, in reliance on a spatial and angular domain channel model respectively, user activity detection and channel estimation algorithms were developed for orthogonal frequency division multiplexing (OFDM) massive MIMO systems in \cite{mke2020} to achieve considerably improved access performance. Beyond those, low-complexity JADCE algorithms were proposed based on dimension reduction \cite{xshao2019} and deep learning \cite{ycui2021}.

While the aforementioned JADCE algorithms exploit the sparsity pattern in the received pilot signal, a common sparsity pattern inherently embedded in both the received pilot and data signal could be further utilized to enhance the performance. The benefits of exploiting such a common sparsity pattern was first witnessed in \cite{ydu2018} for massive RA systems with a single-antenna BS, where a joint activity detection, channel estimation and MUD algorithm was proposed under the framework of MMV CS. An extension for multi-antenna BSs was conducted in \cite{qzou2020} via the bilinear generalized AMP (BiG-AMP) algorithm \cite{jtpa2014}. These preliminary attempts, however, were limited to uncoded transmissions and failed to take advantages of channel coding in modern digital communication systems. Specifically, the error detection mechanism of channel codes can be utilized to determine a subset of active users with high channel quality \cite{xbian2021}. Besides, the soft decoding results, which carry posterior information of the transmitted data symbols, are valuable for improving the accuracy of activity detection, channel estimation and MUD. However, the joint detection/decoding problem is computationally infeasible for channel codes even with reasonable block lengths \cite{bmho2003}. Thus, it necessitates a holistic investigation of how channel decoders can be effectively integrated with other critical components in a massive RA receiver, including user activity detection, channel estimation, and MUD, which will be pursued in this paper. 

\subsection{Contributions}
In this paper, we endeavor to push the performance limit of uplink receivers for grant-free massive RA by leveraging both the common sparsity pattern and channel decoding results. Our main contributions lie in developing effective methods to tackle the algorithmic and computational challenges of the joint design of activity detection, channel estimation, and data decoding, as summarized below.

\begin{itemize}
    \item We propose a turbo receiver for joint activity detection, channel estimation, and data decoding, which iterates between a joint estimator and a channel decoder. In order to exploit the common sparsity pattern in the received pilot and data signal, the joint estimator for joint activity detection, channel estimation, and data symbol detection, is developed by solving a \emph{bilinear inference} problem based on the BiG-AMP algorithm. To boost its performance, we further leverage the posterior log-likelihood ratios (LLRs) of the data bits from the channel decoder to derive the extrinsic information, which serves as the updated estimates of the user activity as well as data symbol distribution and are used as prior information for the next turbo iteration.
\end{itemize}

\begin{itemize}
    \item Albeit the turbo receiver is effective in exploiting the common sparsity pattern, the BiG-AMP-based joint estimator incurs significant computation overhead. To facilitate fast execution while retaining the performance gain of the iterative receiver, we develop a side information (SI)-aided receiver that executes a sequential estimator and a channel decoder alternatively. The sequential estimator is developed for JADCE based on the AMP algorithm, which processes only the received pilot signal, leaving data symbol detection to a minimum mean square error (MMSE) equalizer. To effectively leverage the common sparsity and channel decoding results, the estimates on \emph{whether a user is active or not} are used as SI for the sequential estimator, which are derived according to the parity check results and posterior LLRs.
\end{itemize}

\begin{itemize}
    \item Simulation results show that the turbo receiver significantly reduces the activity detection, channel estimation, and data decoding errors compared with the baseline schemes. Remarkably, in the simulated setting, assuming the block error rate (BLER) requirement is $10^{-3}$, the turbo receiver is able to support 40 active users while the separate design can only support 20. Meanwhile, the SI-aided receiver saves more than 60\% of the execution time compared with the turbo receiver, while maintaining a noticeable performance improvement compared against a data-assisted design that only leverages the common sparsity pattern.
\end{itemize}

\subsection{Organization}
The rest of this paper is organized as follows. We introduce the system model in Section \uppercase\expandafter{\romannumeral2}. A turbo receiver for joint activity detection, channel estimation, and data decoding is developed in Section \uppercase\expandafter{\romannumeral3}. In Section \uppercase\expandafter{\romannumeral4}, we propose a low-complexity SI-aided receiver. Simulation results are presented in Section \uppercase\expandafter{\romannumeral5}, and Section \uppercase\expandafter{\romannumeral6} concludes this paper.

\subsection{Notations} 
We use lower-case letters, bold-face lower-case letters, bold-face upper-case letters, and math calligraphy letters to denote scalars, vectors, matrices, and sets, respectively. The entry in the $i$-th row and $j$-th column of matrix $\mathbf{M}$ is denoted as $m_{ij}$, and the matrix transpose, complex conjugate, and conjugate transpose operators are denoted as $(\cdot)^{\mathrm{T}}$, $(\cdot)^{\mathrm{*}}$, and $(\cdot)^{\mathrm{H}}$, respectively. Besides, $\mathbf{M}_{\backslash i, j}$ represents all the elements in matrix $\mathbf{M}\triangleq[m_{ij}]$ except $m_{ij}$. In addition, $\exp\left(\cdot\right)$ denotes the exponential function, $\delta\left(\cdot\right)$ denotes the Dirac delta function, $\lfloor \cdot \rfloor$ denotes the floor function, and $\mathcal{CN}(x;\mu,v)$ denotes the probability density function (PDF) of a complex Gaussian random variable $x$ with mean $\mu$ and variance $v$.

\section{System Model}
\begin{figure}[t]
\centering
\includegraphics[width=3.2in]{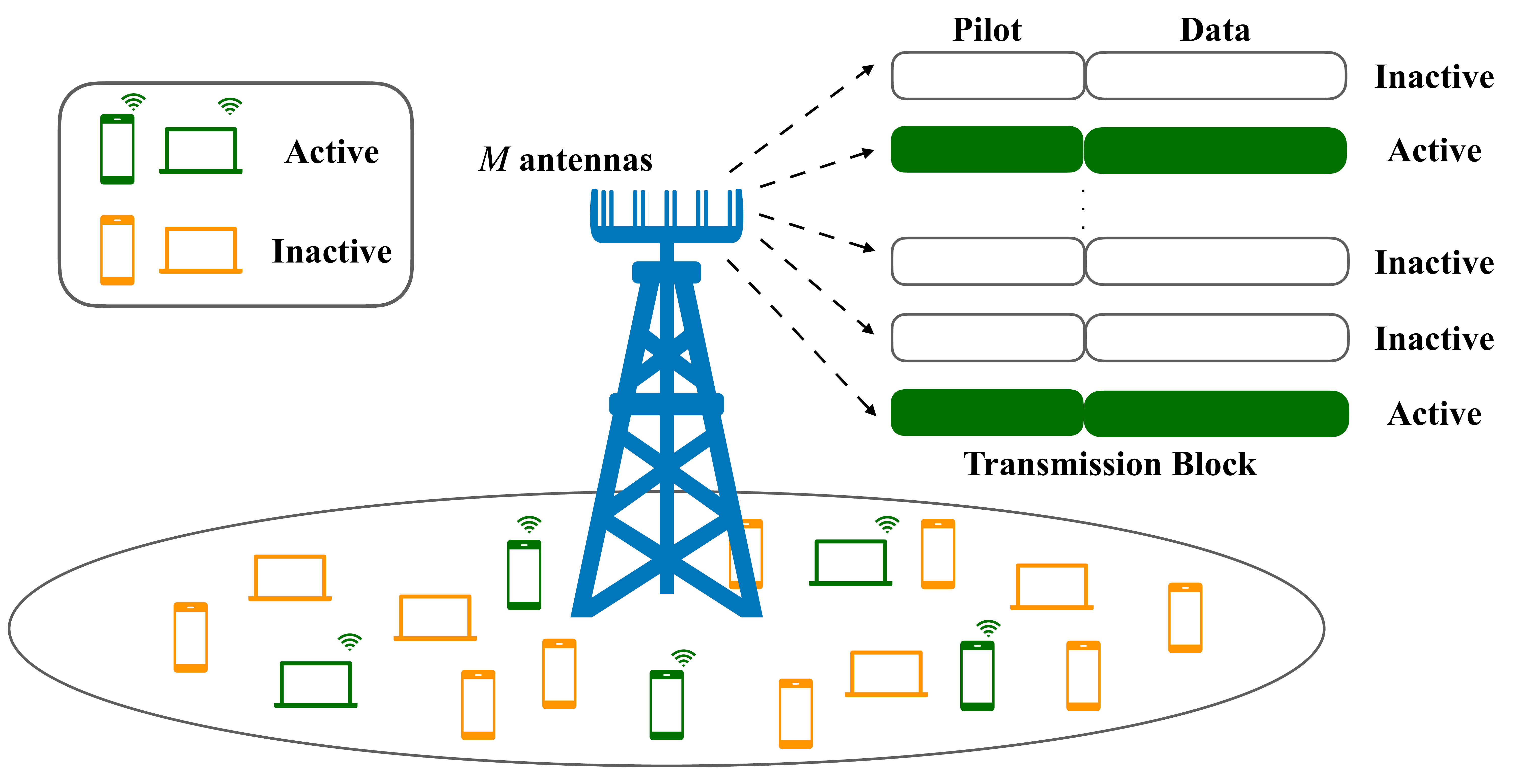}
\caption{System model and the transmission block structure.}
\label{systemmodel}
\end{figure}

We consider an uplink cellular system as shown in Fig. \ref{systemmodel}, where a BS with $M$ antennas serves $N$ single-antenna users. The users are assumed to have short data packets to transmit occasionally, and at each time instant, $K$ ($K\leq N$) among the $N$ users become active for transmission. Denote $u_{n}\in\{0,1\}$ as the user activity indicator, where $u_{n}=1$ means user $n$ is active and $u_{n} = 0$ if it is inactive. The sets of system users and active users are represented by $\mathcal{N}\triangleq \{1,\cdots, N\}$ and $\Xi \triangleq \left\{n \in \mathcal{N} | u_{n}=1 \right\}$, respectively, and the set of BS antennas is denoted as $\mathcal{M}\triangleq\{1,\cdots,M\}$. Besides, the number of BS antennas is assumed to be no less than the number of active users, i.e.,  $M\geq K$, to avoid the system from being overloaded \cite{kkwong2007}.

Each transmission block contains $T$ symbol intervals and we assume quasi-static block fading channels, i.e., the channel state remains unchanged within a transmission block, but varies independently across multiple blocks. The uplink channel vector from user $n$ to the BS is modeled as $\mathbf{f}_{n}=\sqrt{\beta_{n}} \bm{\alpha}_{n}$, where $\bm{\alpha}_{n}$ and $\beta_{n}$ denote the small-scale and large-scale fading coefficients, respectively. We focus on Rayleigh fading channels, i.e., $\bm{\alpha}_{n} \sim \mathcal{CN}(\bm{0},\mathbf{I}_{M})$\footnote{With slight abuse of notation, $\mathcal{CN}\left(\bm{\mu},\bm{\Sigma}\right)$ also denotes the complex Gaussian distribution with mean $\bm{\mu}$ and covariance matrix $\bm{\Sigma}$.}, and assume that the users are static with $\{\beta_{n}\}$'s known by the BS \cite{zchen2018}.

A grant-free RA scheme where each transmission block is divided into two phases, as shown in Fig. \ref{systemmodel}, is adopted for uplink transmission. Specifically, in the first phase, $L$ symbols, denoted as $\mathbf{T}_{p}$, are reserved for pilot transmission, which are essential for user activity detection and channel estimation at the BS; Whereas the remaining $L_{d}\triangleq T-L$ symbols, denoted as $\mathbf{T}_{d}$, are used for payload delivery in the second phase. It is important to note that although orthogonal pilot sequences are most beneficial for accurate channel estimation, it is strictly prohibitive in mMTC since the number of users can be much larger than the pilot length, i.e., $N \gg L$. As a result, we assign the users with a set of non-orthogonal and unique pilot sequences $\{\mathbf{x}_{p n}\}$'s by sampling a complex Gaussian distribution, i.e., $\mathbf{x}_{p n}\triangleq \left[x_{n 1}, \cdots, x_{n L}\right]$ with $x_{nl} \sim \mathcal{C} \mathcal{N}\left(0, 1\right)$, which achieves asymptotic orthogonality when $L$ is sufficiently large. Define $\mathbf{X}_{p}\triangleq \left[\mathbf{x}_{p 1}, \cdots, \mathbf{x}_{p N}\right]^{\mathrm{T}}$ as the collection of pilot sequences. 

In each transmission block, $N_{b}$ payload bits, denoted as $\bm{b}_{n}\triangleq \left[b_{n 1}, \cdots, b_{n N_{b}}\right], n \in \Xi$, need to be transmitted for each active user, which are encoded for error detection and correction. Following contemporary communication standards such as the long term evolution (LTE) \cite{3gpp2011} and 5G new radio (NR) \cite{3gpp2020}, cyclic redundancy check (CRC) bits are generated and attached to the payload bits to form a code block. We represent the CRC generation and attachment procedures by function $\Upsilon$$:\{0,1\}^{N_{b}} \rightarrow \{0,1\}^{N_d}$, where $N_{d}$ denotes the size of a code block. Thus, the code blocks of the active users can be expressed as follows:
\begin{align}
\bm{d}_{n} \triangleq \left[d_{n 1}, \cdots, d_{n N_{d}}\right]=\Upsilon(\bm{b}_{n}), n \in \Xi.
\end{align}

\noindent Each code block is then encoded by a channel encoder, which is abstracted as function $\Phi$$:\{0,1\}^{N_{d}} \rightarrow \{0,1\}^{N_c}$, and the coded bits can be represented as follows:
\begin{align}
\bm{c}_{n} \triangleq \left[c_{n 1}, \cdots, c_{n N_{c}}\right]=\Phi(\bm{d}_{n}), n \in \Xi.
\end{align}

\noindent Note that $N_c$ is the number of coded bits and the code rate $\phi$ is defined as the ratio between $N_d$ and $N_c$, i.e., $\phi \triangleq \frac{N_{d}}{N_{c}}$.

The coded bits are modulated to a set of constellation points $\mathcal{X}$ with normalized average power via an invertible mapping $\mu$$:\{0,1\}^{\log_{2}|\mathcal{X}|}\rightarrow \mathcal{X}$, i.e., for an arbitrary bit sequence with length $\log_2 |\mathcal{X}|$, $\mu\left([c_{1},\cdots,c_{\log_{2}|\mathcal{X}|}] \right) = s$ if and only if $\mu^{-1}\left(s\right) = \left[c_{1},\cdots,c_{\log_{2}|\mathcal{X}|}\right]$, where $s\in\mathcal{X}$ is a constellation point.
The modulated symbols for the active users are denoted as follows:
\begin{align}
\mathbf{x}_{d n}\triangleq \left[x_{n (L+1)}, \cdots, x_{n T}\right], n \in \Xi.
\end{align}

\noindent We assume $N_{c}= L_{d}\log_2|\mathcal{X}|$ for simplicity and assign zero vectors to $\mathbf{x}_{dn}$ for the set of inactive users. Let $\mathbf{X}_{d}\triangleq \left[\mathbf{x}_{d 1}, \cdots, \mathbf{x}_{d N}\right]^{\mathrm{T}}$ denote the transmitted data symbols from all the users. As a result, the received signal of the transmission block $\tilde{\mathbf{Y}} \in \mathbb{C}^{M\times T}$ at the BS can be expressed as follows:
\begin{align}
\tilde{\mathbf{Y}}\triangleq \left[\tilde{\mathbf{Y}}_{p}, \tilde{\mathbf{Y}}_{d}\right]=\sqrt{\gamma}\mathbf{H}\underbrace{\left[\mathbf{X}_{p},\mathbf{X}_{d}\right]}_{\triangleq \mathbf{X}}+\underbrace{\left[\tilde{\mathbf{N}}_{p},\tilde{\mathbf{N}}_{d}\right]}_{\triangleq \tilde{\mathbf{N}}},
\end{align}

\noindent where $\gamma$ is the uplink transmit power, $\tilde{\mathbf{Y}}_{p} \in \mathbb{C}^{M\times L}$ and $\tilde{\mathbf{Y}}_{d} \in \mathbb{C}^{M\times L_{d}}$ are the received pilot and data signal, respectively, and $\mathbf{H} \triangleq \left[\mathbf{h}_{1},...,\mathbf{h}_{N}\right] \in \mathbb{C}^{M\times N}$ with $\mathbf{h}_{n}\triangleq u_{n}\mathbf{f}_{n}$ denotes the \emph{effective channel coefficient} matrix. Besides, $\tilde{\mathbf{N}}=\left[\tilde{\mathbf{n}}_{1},...,\tilde{\mathbf{n}}_{T}\right]\in \mathbb{C}^{M\times T}$ is the additive white Gaussian noise (AWGN) with zero mean and variance $\sigma^{2}$ for each element, and $\tilde{\mathbf{N}}_{p} \in \mathbb{C}^{M\times L}$ and $\tilde{\mathbf{N}}_{d} \in \mathbb{C}^{M\times L_{d}}$ are the noise of the received pilot and data signal, respectively. The noise variance $\sigma^2$ is assumed known, which can be accurately estimated at the BS \cite{tcui2006}. Define $\mathbf{Y}\triangleq \tilde{\mathbf{Y}} \slash \sqrt{\gamma}$, $\mathbf{Y}_{p}\triangleq \tilde{\mathbf{Y}}_{p} \slash \sqrt{\gamma}$, $\mathbf{Y}_{d} \triangleq \tilde{\mathbf{Y}}_{d} \slash \sqrt{\gamma}$, $\mathbf{N}\triangleq \tilde{\mathbf{N}} \slash \sqrt{\gamma}$, $\mathbf{N}_{p}\triangleq \tilde{\mathbf{N}}_{p} \slash \sqrt{\gamma}$, and $\mathbf{N}_{d}\triangleq \tilde{\mathbf{N}}_{d} \slash \sqrt{\gamma}$ as the normalized received signals and noise for the ease of notation. TABLE \ref{Def} summarizes the key notations in this paper and their definitions.
\begin{table}[ht]
\caption{Key Notations and Their Definitions}
\centering
\scalebox{0.95}{
    \begin{tabular}{c|c}
    \hline
     Notation & Definition  \\
    \hline
    $M$, $N$, $K$ & Number of BS antennas, system users, and active users\\
    \hline
    $\mathcal{M}$, $\mathcal{N}$, $\Xi$ & Set of BS antennas, system users, and active users\\
    \hline
    $T$, $L$, $L_{d}$ & \makecell{Number of symbols, pilot symbols, \\and data symbols in a transmission block}\\
    \hline
    $\bm{\alpha}_{n}$, $\beta_{n}$ & Small-scale and large-scale fading coefficients \\
    \hline
    $u_{n}$ & User activity indicator \\
    \hline
    $\mathbf{h}_{n}$, $\mathbf{H}$ & Effective channel coefficient vector and matrix \\
    \hline
    $\bm{b}_{n}$, $\bm{d}_{n}$, $\bm{c}_{n}$ & Payload bits, code block, and coded bits \\
    \hline
    $N_{b}$, $N_{d}$, $N_{c}$ & Number of payload bits, code block bits, and coded bits\\
    \hline
    $\mathcal{X}$ & Set of constellation points \\
    \hline
    $\mathbf{X}_{p}$, $\mathbf{X}_{d}$& Pilot sequences and modulated data symbols \\
    \hline
    $\tilde{\mathbf{Y}}_{p}$, $\tilde{\mathbf{Y}}_{d}$ & Received pilot and data signals \\
    \hline
    $\mathbf{Y}_{p}$, $\mathbf{Y}_{d}$ & Normalized received pilot and data signals\\
    \hline
    $\gamma$, $\sigma^2$ & Uplink transmit power and noise variance \\
    \hline
    $\hat{\Xi}$, $\hat{\Xi}_{c}$ & Estimated set of active users and users that pass CRC\\
    \hline
    $\hat{\bm{d}}_{n}$, $\hat{\bm{b}}_{n}$ & Detected code blocks and payload bits \\
    \hline
    \end{tabular}}
\label{Def}
\end{table}

In the following sections, we will develop efficient algorithms to detect the set of active users, estimate their channel coefficients and the transmitted payload bits.

\section{Joint Estimation via a Turbo Receiver}
In this section, we develop a turbo receiver for joint estimation of the user activity, channel coefficients, and payload data of the active users. It is noteworthy that while the turbo principle has achieved great success in conventional multi-user MIMO systems \cite{shay2004,xwa2004}, its applications in grant-free massive RA are still unchartered due to the new requirement of user activity detection. Besides, in contrast to most state-of-the-art approaches that follow a sequential user activity detection and data detection/decoding pipeline \cite{lliu2018}, our design exploits the common sparsity pattern in both the received pilot and data signal. Meanwhile, it takes advantages of the soft decoding information in order to optimize the activity detection and data reception performance.
\subsection{Overview of the Turbo Receiver}
\begin{figure}[t]
\centering
\includegraphics[width=3.4in]{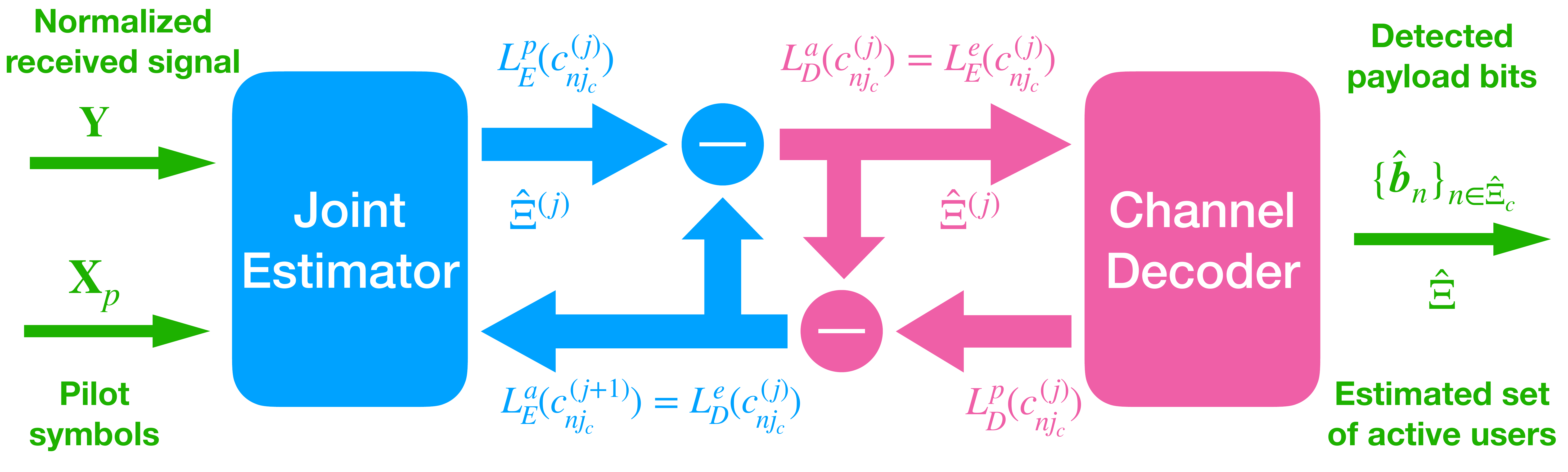}
\caption{The proposed turbo receiver for massive RA.}
\label{jointmodel}
\end{figure}

The proposed turbo receiver iterates between a joint estimator and a channel decoder as shown in Fig. \ref{jointmodel}, which is inspired by the \emph{turbo decoding principle} \cite{cbe1996} that leverages multiple concatenated elementary decoders with the aid of the extrinsic information. In particular, responsible for user activity detection, channel estimation, and soft data symbol detection, the joint estimator is designed based on the BiG-AMP algorithm \cite{jtpa2014}. It also estimates the posterior probabilities of the transmitted data symbols in each turbo iteration, which are converted as extrinsic information of the coded bits. On the other hand, the channel decoder is developed based on the belief propagation (BP) algorithm \cite{frks2001}, which accepts the extrinsic information of the coded bits as input to generate their posteriors. The extrinsic LLRs of the coded bits, i.e., the logarithm of ratio between the probabilities that a coded bit is “0” or “1”, are obtained accordingly and translated to priors of the transmitted data symbols for the use of the joint estimator in the next turbo iteration. The turbo iteration terminates after $Q_{1}$ rounds or when an exit condition is achieved, after which, hard decision is performed to obtain the code block, followed by a cyclic redundancy check. The workflow of the turbo receiver is summarized in Algorithm 1 with details of the joint estimator and channel decoder to be elaborated in the following subsections. Note that an initial estimate of the effective channel coefficients and their variances, as well as the \emph{average sparsity levels} derived from the estimated effective channel coefficients, are obtained via the AMP algorithm developed in \cite{mke2020}\footnote{With prior knowledge of the user active probability, the AMP algorithm \cite{mke2020} estimates the effective channel coefficients and their variances based on the received pilot signal, and a set of belief indicators $\{\tilde{\rho}_{mn}\}$'s are derived as the posterior probabilities of the effective channel coefficients to be non-zero. In this paper, we term $\tilde{\rho}_{mn}$ as the \emph{posterior sparsity level} of user $n$ at the $m$-th BS antenna, and define $\bar{\rho}_{n} \triangleq \frac{1}{M}\sum_{m\in\mathcal{M}}\tilde{\rho}_{mn}$ as the \emph{average sparsity level} of user $n$, which is a reliable statistic of the activity status and updated iteratively by the joint estimator of the proposed turbo receiver.}.

\begin{algorithm}[htpb]
\caption{The Proposed Turbo Receiver for Massive RA}
{\bf Input:}
The normalized received signal $\mathbf{Y}$, pilot symbols $\mathbf{X}_{p}$, maximum number of iterations $Q_{1}$, and accuracy tolerance $\epsilon_{1}$.\\
{\bf Output:}
The estimated set of active users $\hat{\Xi}$, the set of users that pass CRC $\hat{\Xi}_{c}$ and their detected payload bits $\{\hat{\bm{b}}_{n}\}$'s.\\
{\bf Initialize:}
$j \leftarrow 0$, $n \in \mathcal{N}$, $\hat{x}_{nt}^{(0)} \leftarrow 0$, $t \in \mathbf{T}_{d}$, $\lambda_{n}^{(0)}=\frac{K}{N}$, $n \in \mathcal{N}$, $\eta_{nt,s}^{(1)} \leftarrow \frac{1}{|\mathcal{X}|}$, $t \in \mathbf{T}_{d}$, $L_{E}^{a}\left(c_{nj_{c}}^{(1)}\right) \leftarrow 0$, $j_{c}=1,\cdots,N_{c}$.
\begin{algorithmic}[1]
\STATE Execute the AMP algorithm in \cite{mke2020} to obtain the initial estimates of the effective channel coefficients $\{\hat{h}_{mn}^{(0)}\}$'s and their variances $\{V_{mn}^{h(0)}\}$'s, and the average sparsity levels $\{\bar{\rho}_{n}^{(0)}\}$'s.
\WHILE{{$j < Q_{1}$} \text{and} {$\frac{\Sigma_{n,t}|\hat{x}_{n t}^{(j)}-\hat{x}_{n t}^{(j-1)}|^{2}}{\Sigma_{n,t}|\hat{x}_{n t}^{(j-1)}|^{2}} > \epsilon_{1}$}}
\STATE $j \leftarrow j+1$
\Statex \quad //\textit{The Joint Estimator}//
\STATE Based on $\{\hat{h}_{mn}^{(j-1)}\}$'s, $\{V_{mn}^{h(j-1)}\}$'s, and $\{\bar{\rho}_{n}^{(j-1)}\}$'s, the
\Statex \quad joint estimator executes Algorithm 2 to estimate the set
\Statex \quad of active users $\hat{\Xi}^{(j)}$, the posterior probabilities that $x_{nt}$ 
\Statex \quad equals $s$, i.e., $\tilde{\eta}_{nt,s}^{(j)}$, $n \in \hat{\Xi}^{(j)}$, $t \in \mathbf{T}_{d}$, and the soft data 
\Statex \quad symbols $\hat{x}_{nt}^{(j)}$, $t \in \mathbf{T}_{d}$.
\STATE Convert $\tilde{\eta}_{nt,s}^{(j)}$ to the posterior LLRs of coded bits 
\Statex \quad $L_{E}^{p}\left(c_{nj_{c}}^{(j)}\right)$, $n \in \hat{\Xi}^{(j)}$ according to (21).
\STATE Calculate the extrinsic information $L_{E}^{e}\left(c_{nj_{c}}^{(j)}\right)$, $n \in \hat{\Xi}^{(j)}$
\Statex \quad as input of the channel decoder $L_{D}^{a}\left(c_{nj_{c}}^{(j)}\right)$, $n \in \hat{\Xi}^{(j)}$ 
\Statex \quad according to (22).
\Statex \quad //\textit{The Channel Decoder}//
\STATE Perform soft data decoding via a BP-based channel
\Statex \quad decoder and obtain the posterior LLRs of the coded bits \Statex \quad $L_{D}^{p}\left(c_{nj_{c}}^{(j)}\right)$, $n \in \hat{\Xi}^{(j)}$.
\STATE Calculate the extrinsic information $L_{D}^{e}\left(c_{nj_{c}}^{(j)}\right)$, $n \in \hat{\Xi}^{(j)}$
\Statex \quad via (24) as input of joint estimator $L_{E}^{a}\left(c_{nj_{c}}^{(j+1)}\right)$ for the
\Statex \quad next turbo iteration, and obtain the prior probabilities
\Statex \quad that $x_{nt}$ equals $s$, i.e., ${\eta}_{nt,s}^{(j+1)}$, $t \in \mathbf{T}_{d}$ according to (26).

\ENDWHILE
\STATE Determine the set of active users $\hat{\Xi}$ as $\hat{\Xi}^{(j)}$.
\STATE Perform hard decision based on $L_{D}^{p}\left(c_{nj_{c}}^{(j)}\right)$ via (27) to obtain the code blocks $\hat{\bm{d}}_{n}$, $n \in \hat{\Xi}$.
\STATE Perform CRC to determine $\hat{\Xi}_{c}$ and detach the CRC bits from $\hat{\bm{d}}_{n}$ to obtain $\hat{\bm{b}}_{n}$, $n \in \hat{\Xi}_{c}$.
\end{algorithmic}
\end{algorithm}
\vspace{-0.4cm}
\subsection{The Joint Estimator}
The joint estimator is designed to detect the set of active users, estimate their channel coefficients and the transmitted data symbols. Since the user activity pattern is encapsulated in $\mathbf{H}$ and can be determined accordingly, it remains for the joint estimator to estimate the effective channel coefficients and soft data symbols. We resort to the MMSE estimators, which can be expressed for the effective channel coefficients and soft data symbols respectively as follows \cite{smkay1993}:
\begin{align}
\hat{h}_{mn} \triangleq \mathbb{E}\left[h_{mn}|\mathbf{Y}\right] = \int h_{mn} p(h_{mn}|\mathbf{Y}) dh_{mn}, \forall m \in \mathcal{M}, n \in \mathcal{N},
\end{align}
\begin{align}
\hat{x}_{nt}\triangleq \mathbb{E}\left[x_{nt}|\mathbf{Y}\right]=\sum x_{nt} p(x_{nt}|\mathbf{Y}), \forall n \in \mathcal{N}, t\in \mathbf{T}_{d},
\end{align}

\noindent where $\hat{h}_{mn}$ ($\hat{x}_{nt}$) is the estimate of $h_{mn}$ ($x_{nt}$), and $p(h_{mn}|\mathbf{Y})$ ($p(x_{nt}|\mathbf{Y})$) denotes the marginal posterior distribution of $h_{mn}$ ($x_{nt}$) given the normalized received signal $\mathbf{Y}$. The marginal posterior distributions can be rewritten in terms of the joint posterior distribution $p(\mathbf{H},\mathbf{X}|\mathbf{Y})$ as follows:
\begin{align}
p\left(h_{mn}|\mathbf{Y}\right)=\int_{\mathbf{H}_{\backslash m, n}} \sum_{\mathbf{X}} p(\mathbf{H},\mathbf{X}|\mathbf{Y})d\mathbf{H},
\end{align}
\begin{align}
p\left(x_{nt}|\mathbf{Y}\right)=\sum_{\mathbf{X}_{\backslash n, t}} \int_{\mathbf{H}} p(\mathbf{H},\mathbf{X}|\mathbf{Y})d\mathbf{H},
\end{align}

\noindent where $p\left(\mathbf{H},\mathbf{X}|\mathbf{Y}\right)$ can be factorized via the Bayes' rule:
\begin{align}
\begin{split}
    p(\mathbf{H},\mathbf{X}|\mathbf{Y})&=\frac{p(\mathbf{Y}|\mathbf{H}, \mathbf{X})p(\mathbf{H})p(\mathbf{X})}{p(\mathbf{Y})}\\
    &\overset{(a)}{=}\frac{1}{p(\mathbf{Y})} p(\mathbf{Y}|\mathbf{H}, \mathbf{X}) p(\mathbf{H|U})p(\mathbf{U}) p(\mathbf{X})\\
    &\overset{(b)}{=}\frac{1}{p(\mathbf{Y})} \prod_{m=1}^{M}\prod_{t=1}^{T} p\left(y_{m t}|\sum_{n=1}^{N} h_{m n} x_{n t}\right) \\
    &\times \prod_{n=1}^{N}\left[p\left(u_{n}\right) \prod_{m=1}^{M} p\left(h_{m n}|u_{n}\right) \prod_{t=1}^{T} p\left(x_{n t}\right)\right].
\end{split}
\end{align}

\noindent In (9), (a) holds since $p(\mathbf{H})=p(\mathbf{H,U})=p(\mathbf{H|U})p(\mathbf{U})$, as the user activity pattern is deterministic given $\mathbf{H}$, and (b) is attributed to the conditional independence of random variables.

\begin{figure}[t]
\centering
\includegraphics[width=3.4in]{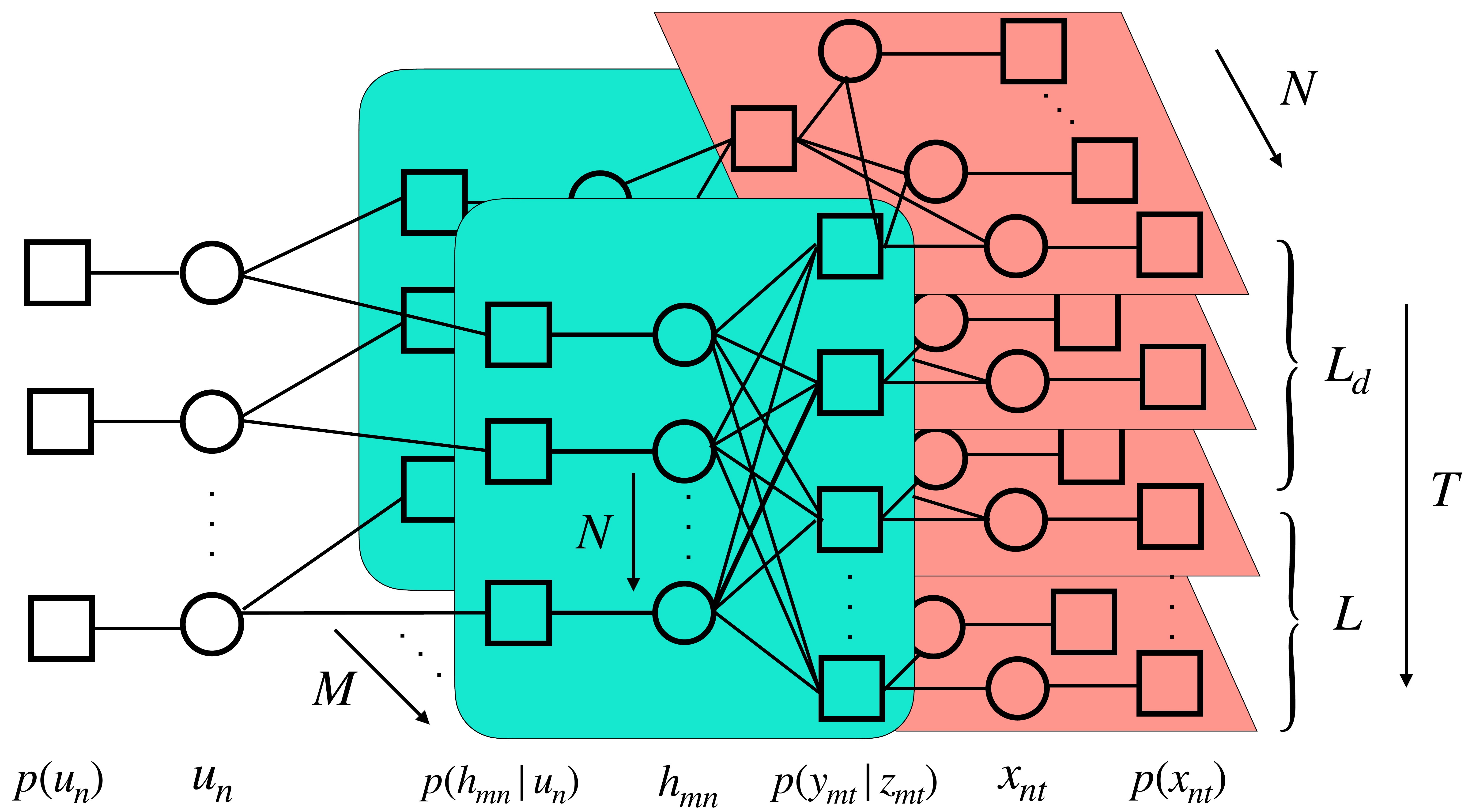}
\caption{The factor graph of the joint posterior distribution $p(\mathbf{H,X|Y})$, where $z_{mt}\triangleq \sum_{n\in\mathcal{N}} h_{m n} x_{n t}$.}
\label{factorgraph}
\end{figure}

Nevertheless, the marginal distributions in (7) and (8) are intractable due to high-dimensional integrals and summations. Fortunately, the factorization in (9) implies efficient approximations via the BP algorithm operating on factor graphs \cite{frks2001}. As shown in Fig. \ref{factorgraph}, a factor graph consists of variable nodes (as indicated by circles), factor nodes (correspond to PDFs as indicated by squares), and edges connecting variable nodes and factor nodes. In the formats of PDFs, messages are propagated in the factor graph and updated iteratively. Specifically, the message from a variable node to a factor node is the product of messages from other adjacent factor nodes of that variable node, while the message from a factor node to a variable node is the integral of the product of that factor and messages from other adjacent variable nodes of that factor node. The posterior PDF of a variable is approximated by the belief of the corresponding variable node, which is the product of messages from all its adjacent factor nodes. For instance, let $I_{x_{nt} \rightarrow f_{y_{mt}}}$ and $I_{f_{y_{mt}} \rightarrow x_{nt}}$ be the messages from variable node $x_{nt}$ to factor node $p(y_{mt}|\sum_{n\in\mathcal{N}} h_{m n} x_{n t})$ and from factor node $p(y_{mt}|\sum_{n\in\mathcal{N}} h_{m n} x_{n t})$ to variable node $x_{nt}$, respectively. They are updated in each iteration of the BP algorithm as follows:
\begin{align}
I_{x_{nt} \rightarrow f_{y_{mt}}} \leftarrow I_{f_{x_{nt}} \rightarrow x_{n t}} \prod_{k\in\mathcal{M}\setminus \{m\}} I_{f_{y_{kt}} \rightarrow x_{n t}},
\end{align}
\begin{align}
\begin{split}
I_{f_{y_{mt}} \rightarrow x_{nt}} &\leftarrow \int p\left(y_{m t} \mid \sum_{k=1}^{N} h_{m k} x_{k t}\right)\\
&\times \prod_{r\in\mathcal{N}\setminus \{n\}} \left(I_{x_{rt} \rightarrow f_{y_{mt}}}\right) \prod_{k\in\mathcal{N}} \left(I_{h_{m k} \rightarrow f_{y_{mt}}}\right) d\mathbf{h}_{m}d\mathbf{x}_{t \backslash n},
\end{split}
\end{align}

\noindent where $I_{f_{x_{nt}} \rightarrow x_{n t}}$ denotes the message from factor node $p(x_{nt})$ to variable node $x_{nt}$ that is used to approximate the prior distribution of $x_{nt}$, and $I_{h_{m n} \rightarrow f_{y_{mt}}}$ is the message from variable node $h_{mn}$ to factor node $p(y_{mt}|\sum_{n\in\mathcal{N}} h_{m n} x_{n t})$. The belief of variable node $x_{nt}$ and the approximated posterior distribution of $x_{nt}$, i.e., $B_{x_{nt}}$ and $r_{x_{nt}}$, are updated via the following expressions, respectively:
\begin{align}
B_{x_{nt}}\leftarrow I_{f_{x_{nt}} \rightarrow x_{n t}} \prod_{m\in\mathcal{M}} I_{f_{y_{mt}} \rightarrow x_{n t}},
\end{align}
\begin{align}
r_{x_{nt}}\leftarrow \frac{B_{x_{nt}}}{\int B_{x_{nt}} dx_{nt}}.
\end{align}

Unfortunately, although the BP algorithm is efficient in calculating marginal distributions, computations of the high-dimensional integrals in (11) still exhibit excessive complexity since $N$ is very large in massive RA systems. To develop a joint estimator with affordable complexity, we turn to the framework of AMP, which is a variant of BP that provides more tractable approximations for marginal distributions \cite{donoho2010}. It adopts the Central Limit Theorem to approximate the product of some messages as a complex Gaussian distribution so that only the mean and variance need to be propagated. Besides, high-order terms of messages are omitted in deriving the means and variances to further reduce the computation complexity. Since the joint estimation of effective channel coefficients and soft data symbols belongs to a bilinear inference problem, the BiG-AMP algorithm \cite{jtpa2014} offers a viable solution by estimating $\mathbf{H}$ and $\mathbf{X}$ alternatively. Key steps of the BiG-AMP-based joint estimator are summarized in Algorithm 2, which is an iterative algorithm that estimates three sets of variables in each iteration, including: 1) The linear mixing variables $\{z_{mt}\}$'s ($z_{mt}\triangleq\sum_{n\in\mathcal{N}} h_{m n} x_{n t}$); 2) The effective channel coefficients $\{h_{mn}\}$'s; and 3) The soft data symbols $\{x_{nt}\}$'s, as elaborated in the following.
\begin{algorithm}[htbp]
\caption{The Joint Estimator based on BiG-AMP}
{\bf Input:}
The normalized received signal $\mathbf{Y}$, pilot symbols $\mathbf{X}_{p}$, the estimates of the likelihood that each user is active $\{\lambda_{n}^{(j)}\}$'s, the estimates of the effective channel coefficients $\{\hat{h}_{mn}^{(j-1)}\}$'s and their variances $\{V_{mn}^{h(j-1)}\}$'s, the prior probabilities that $x_{nt}$ equals $s$, i.e., $\{{\eta}_{nt,s}^{(j)}\}$'s, the threshold of determining the active user $\theta$, maximum number of iterations $Q_{2}$, and accuracy tolerance $\epsilon_{2}$.\\
{\bf Output:}
The estimated set of active users $\hat{\Xi}^{(j)}$, and the posterior probabilities that $x_{nt}$ equals $s$, i.e., $\tilde{\eta}_{nt,s}^{(j)}$, $n \in \hat{\Xi}^{(j)}$, $t \in \mathbf{T}_{d}$.\\
{\bf Initialize:}
$i \leftarrow 0$, $\hat{h}_{mn}^{(j)}(0)\leftarrow \hat{h}_{mn}^{(j-1)}$, $V_{mn}^{h(j)}(0)\leftarrow V_{mn}^{h(j-1)}$, $\hat{s}_{mt}^{(j)}(0)\leftarrow 0$, $\hat{x}_{nt}^{(j)}(0)\leftarrow 0$, $t \in \mathbf{T}_{d}$, $V_{nt}^{x(j)}(0)\leftarrow 1$, $t \in \mathbf{T}_{d}$.
\begin{algorithmic}[1]
\WHILE{{$i < Q_{2}$} \text{and} {$\frac{\Sigma_{m,t}|\hat{z}_{m t}^{(j)}(i)-\hat{z}_{m t}^{(j)}(i-1)|^{2}}{\Sigma_{m,t}|\hat{z}_{m t}^{(j)}(i-1)|^{2}} > \epsilon_{2}$}}
\STATE $i\gets i+1$
\Statex \quad //\textit{Estimate the Linear Mixing Variable ${z}_{m t}$}//

\STATE $\forall m,t: M_{m t}^{p(j)}(i)=\sum_{n} \hat{h}_{m n}^{(j)}(i-1) \hat{x}_{n t}^{(j)}(i-1)-\hat{s}_{m t}^{(j)}(i-1)$

\Statex \quad $\sum_{n}\Big(|\hat{x}_{n t}^{(j)}(i-1)|^{2} V_{m n}^{h(j)}(i-1)+|\hat{h}_{m n}^{(j)}(i-1)|^{2}V_{n t}^{x(j)}(i-1)\Big)$

\STATE $\forall m,t: V_{m t}^{p(j)}(i)=\sum_{n}\Big(|\hat{x}_{n t}^{(j)}(i-1)|^{2} V_{m n}^{h(j)}(i-1)$

\Statex \quad $+|\hat{h}_{m n}^{(j)}(i-1)|^{2}V_{n t}^{x(j)}(i-1)\Big)+\sum_{n} V_{m n}^{h(j)}(i-1)V_{n t}^{x(j)}(i-1)$

\STATE $\forall m,t: \hat{z}_{m t}^{(j)}(i)=\mathbb{E}\left[z_{mt}|M_{m t}^{p(j)}(i),V_{m t}^{p(j)}(i)\right]$

\Statex \quad $=\frac{y_{mt}V_{mt}^{p(j)}(i)+(\sigma^{2}/\gamma)M_{mt}^{p(j)}(i)}{(\sigma^{2}/\gamma)+V_{mt}^{p(j)}(i)}$

\STATE $\forall m, t:V_{m t}^{z(j)}(i)=\text{Var}\left[z_{mt}|M_{m t}^{p(j)}(i),V_{m t}^{p(j)}(i)\right]$

\Statex \quad $=\frac{(\sigma^{2}/\gamma) V_{m t}^{p(j)}(i)}{(\sigma^{2}/\gamma)+V_{m t}^{p(j)}(i)}$

\STATE $\forall m, t: \hat{s}_{m t}^{(j)}(i)=\left(\hat{z}_{m t}^{(j)}(i)-M_{m t}^{p(j)}(i)\right)/V_{m t}^{p(j)}(i)$

\STATE $\forall m, t: V_{m t}^{s(j)}(i)=\left(1-V_{m t}^{z(j)}(i)/{V_{m t}^{p(j)}(i)}\right)/V_{m t}^{p(j)}(i)$

\Statex \quad //\textit{Estimate the Effective Channel Coefficients} //

\STATE $\forall m, n: Q_{p,m n}^{h(j)}(i)=\left(\sum_{t \in \mathbf{T}_{p}}|x_{nt}|^{2}V_{m t}^{s(j)}(i)\right)^{-1}$

\STATE $\forall m, n: P_{p,m n}^{h(j)}(i)=\hat{h}_{mn}^{(j)}(i-1)+Q_{p,m n}^{h(j)}(i)$

\Statex \quad $\sum_{t \in \mathbf{T}_{p}}x_{nt}^{*}\hat{s}_{m t}^{(j)}(i)\cdot$

\STATE $\forall m, n: Q_{d,m n}^{h(j)}(i)=\left(\sum_{t \in \mathbf{T}_{d}}\left|\hat{x}_{nt}^{(j)}(i-1)\right|^{2}V_{m t}^{s(j)}(i)\right)^{-1}$

\STATE $\forall m, n: P_{d,m n}^{h(j)}(i)=\hat{h}_{mn}^{(j)}(i-1)\Big(1-Q_{d,m n}^{h(j)}(i)\sum_{t \in \mathbf{T}_{d}}$

\Statex \quad $V_{n t}^{x(j)}(i-1)V_{m t}^{s(j)}(i)\Big)+Q_{d,m n}^{h(j)}(i)\sum_{t\in \mathbf{T}_{d}}\hat{x}_{nt}^{(j)*}(i-1)\hat{s}_{m t}^{(j)}(i)$

\STATE $\forall m, n: P_{m n}^{h(j)}(i)=\Big(P_{p,m n}^{h(j)}(i)Q_{d,m n}^{h(j)}(i)+P_{d,m n}^{h(j)}(i)\cdot$

\Statex \quad $Q_{p,m n}^{h(j)}(i)\Big)\Big/\left(Q_{p,m n}^{h(j)}(i)+Q_{d,m n}^{h(j)}(i)\right)$

\STATE $\forall m, n: Q_{m n}^{h(j)}(i)=Q_{p,m n}^{h(j)}(i)Q_{d,m n}^{h(j)}(i)\Big/\Big(Q_{p,m n}^{h(j)}(i)$

\Statex \quad $+ Q_{d,m n}^{h(j)}(i)\Big)$

\STATE $\forall m, n: K_{mn}^{(j)}(i)=\ln\left(\frac{\mathcal{CN}\left(0;P_{mn}^{h(j)}(i),Q_{mn}^{h(j)}(i)+\beta_{n}\right)}{\mathcal{CN}\left(0;P_{mn}^{h(j)}(i),Q_{mn}^{h(j)}(i)\right)}\right)$

\Statex \quad $=\ln \left(\frac{Q_{m n}^{h(j)}(i)}{Q_{m n}^{h(j)}(i)+\beta_{n}}\right)+\frac{\left|P_{mn}^{h(j)}(i)\right|^{2}\beta_{n}}{\left(Q_{m n}^{h(j)}(i)+\beta_{n}\right)Q_{m n}^{h(j)}(i)}$
\STATE $\forall m, n: L_{mn}^{(j)}(i)=\ln \left( \frac{\lambda_{n}^{(j)}}{1-\lambda_{n}^{(j)}}\right)+\sum_{k \in \mathcal{M} \backslash \{m\}} \left(K_{kn}^{(j)}(i)\right)$
\STATE $\forall m, n: \rho_{mn}^{(j)}(i)=\exp\left(L_{mn}^{(j)}(i)\right)\Big/\left(1+\exp\left(L_{mn}^{(j)}(i)\right)\right)$
\algstore{myalg}
\end{algorithmic}
\end{algorithm}

\begin{algorithm}              
\begin{algorithmic} [1]            
\algrestore{myalg}
\STATE $\forall m, n: \mu_{mn}^{(j)}(i)=\beta_{n}P_{mn}^{h(j)}(i)\Big/\left(\beta_{n}+Q_{mn}^{h(j)}(i)\right)$, 
\Statex \quad $\tau_{mn}^{(j)}(i)=\beta_{n}Q_{mn}^{h(j)}(i)\Big/\left(\beta_{n}+Q_{mn}^{h(j)}(i)\right)$

\STATE $\forall m, n: \tilde{\rho}_{mn}^{(j)}(i)=\rho_{mn}^{(j)}(i)\Big/\Big(\rho_{mn}^{(j)}(i)+\left(1-\rho_{mn}^{(j)}(i)\right)\cdot$

\Statex \quad $\exp\Big(-\ln\left(\frac{Q_{m n}^{h(j)}(i)}{Q_{m n}^{h(j)}(i)+\beta_{n}}\right)-\frac{|P_{mn}^{h(j)}(i)|^{2}\beta_{n}}{\left(Q_{m n}^{h(j)}(i)+\beta_{n}\right)Q_{m n}^{h(j)}(i)}\Big)\Big)$

\STATE $\forall m, n:\hat{h}_{mn}^{(j)}(i)=\mathbb{E}\left[h_{mn}|P_{m n}^{h(j)}(i),Q_{m n}^{h(j)}(i)\right]$

\Statex \quad $=\tilde{\rho}_{mn}^{(j)}(i)\mu_{mn}^{(j)}(i)$

\STATE $\forall m, n: V_{m n}^{h(j)}(i)=\text{Var}\left[h_{mn}|P_{m n}^{h(j)}(i),Q_{m n}^{h(j)}(i)\right]$

\Statex \quad $=\tilde{\rho}_{mn}^{(j)}(i)\left(\left|\mu_{mn}^{(j)}(i)\right|^{2}+\tau_{mn}^{(j)}(i)\right)-\left|\hat{h}_{mn}^{(j)}(i)\right|^{2}$
\Statex \quad //\textit{Estimate the Soft Data Symbols}//
\STATE $\forall n, t \in \mathbf{T}_{d}: Q_{n t}^{x(j)}(i)=\left(\sum_{m}\left|\hat{h}_{mn}^{(j)}(i-1)\right|^{2}V_{m t}^{s(j)}(i)\right)^{-1}$

\STATE $\forall n, t \in \mathbf{T}_{d}:P_{n t}^{x(j)}(i)=\hat{x}_{nt}^{(j)}(i-1)\Big(1-Q_{n t}^{x(j)}(i)\cdot$

\Statex \quad $\sum_{m}V_{m n}^{h(j)}(i-1)V_{m t}^{s(j)}(i)\Big)+Q_{n t}^{x(j)}(i) \sum_{m}\hat{h}_{mn}^{(j)*}(i-1)\hat{s}_{m t}^{(j)}(i)$

\STATE $\forall n, t \in \mathbf{T}_{d}:\tilde{\eta}_{nt,s}^{(j)}(i)=\frac{\eta_{nt,s}^{(j)}\mathcal{C}\mathcal{N}\left(s;P_{nt}^{x(j)}(i),Q_{nt}^{x(j)}(i)\right)}{\sum_{s'\in \mathcal{X}}\eta_{nt,s'}^{(j)}\mathcal{C}\mathcal{N}\left(s';P_{nt}^{x(j)}(i),Q_{nt}^{x(j)}(i)\right)}$

\STATE $\forall n, t \in \mathbf{T}_{d}:\hat{x}_{nt}^{(j)}(i)=\mathbb{E}\left[x_{nt}|P_{n t}^{x(j)}(i),V_{n t}^{x(j)}(i)\right]$

\Statex \quad $=\bar{\rho}_{n}^{(j-1)}\sum_{s \in \mathcal{X}}\tilde{\eta}_{nt,s}^{(j)}(i)s$

\STATE $\forall n, t \in \mathbf{T}_{d}: V_{n t}^{x(j)}(i)=\text{Var}\left[x_{nt}|P_{n t}^{x(j)}(i),V_{n t}^{x(j)}(i)\right]$

\Statex \quad $=\sum_{s\in \mathcal{X}}\tilde{\eta}_{nt,s}^{(j)}(i)\left|\bar{\rho}_{n}^{(j-1)}s-\hat{x}_{nt}^{(j)}(i)\right|^{2}$

\ENDWHILE
\STATE Update $\bar{\rho}_{n}^{(j)}=\frac{1}{M}\sum_{m \in \mathcal{M}} \tilde{\rho}_{mn}^{(j)}$, $\hat{h}_{mn}^{(j)}$, and $V_{mn}^{h(j)}$ for the next turbo iteration.
\STATE Update $\lambda_{n}^{(j+1)}=\kappa \bar{\rho}_{n}^{(j)}+(1-\kappa)\lambda_{n}^{(j)}$, $\forall n \in \mathcal{N}$.
\STATE Determine the estimated set of active users as $\hat{\Xi}^{(j)} \triangleq \{n \in \mathcal{N} \mid \bar{\rho}_{n}^{(j)} \geq\theta\}$.

\end{algorithmic}
\end{algorithm}

\subsubsection{Estimate the linear mixing variables} In each iteration, the joint estimator first estimates the linear mixing variable $z_{m t}$ from $y_{m t}$. The basic principle is that if the prior distribution of a variable and its likelihood function are available, the posterior probability can be derived for MMSE estimation by using the Bayes' rule. Since $y_{mt}=z_{mt}+n_{mt}$, the likelihood function is given as follows:
\begin{align}
p\left(y_{m t}|\sum_{n\in\mathcal{N}} h_{m n} x_{n t}\right)=\frac{\gamma}{\pi \sigma^{2}}\text{exp}\left(-\frac{\gamma}{\sigma^{2}}\left|y_{m t}-\sum_{n\in\mathcal{N}} h_{m n} x_{n t}\right|^{2}\right).
\end{align}

\noindent The prior distribution of $z_{mt}$ is approximated as a complex Gaussian distribution with mean $M_{m t}^{p(j)}(i)$ and variance $V_{m t}^{p(j)}(i)$ in the $i$-th iteration of the joint estimator, as shown in Line 3 and 4 in Algorithm 2, respectively, where the superscript “$^{(j)}$” denotes the turbo iteration index, $\hat{h}_{mn}^{(j)}(i-1)$ and $V_{mn}^{h(j)}(i-1)$ are the most updated estimate of the effective channel coefficient and its variance, $\hat{x}_{nt}^{(j)}(i-1)$ and $V_{nt}^{x(j)}(i-1)$ are the latest estimate of the soft data symbol and its variance, and $\hat{s}_{mt}^{(j)}(i-1)$ denotes the \emph{scaled residual} of $z_{mt}$. Note that for $t\in \mathbf{T}_{p}$, we have $\hat{x}_{nt}^{(j)}(i-1)=x_{nt}$ and $V_{nt}^{x(j)}(i-1)=0$ as the pilot symbols are known at the BS. 

Since both the approximated prior distribution of $z_{mt}$ and the likelihood function of $z_{mt}$ are complex Gaussian, the posterior distribution $p\left(z_{mt}|y_{mt}\right)$ can also be approximated by a complex Gaussian distribution with mean $\hat{z}_{mt}^{(j)}(i)$ and variance $V_{mt}^{z(j)}(i)$ given in Line 5 and 6 in Algorithm 2, respectively. Detailed derivations are provided in Appendix A. Note that the posterior mean of $z_{mt}$ also gives the MMSE estimate $\hat{z}_{mt}^{(j)}(i)$. Besides, the \emph{scaled residual} $\hat{s}_{mt}^{(j)}(i)$ of $z_{mt}$ and the corresponding \emph{inverse-residual-variance} $V_{m t}^{s(j)}(i)$ are updated in Line 7 and 8, respectively, which are useful for approximating the likelihood functions of the effective channel coefficients and soft data symbols. 

\subsubsection{Estimate the effective channel coefficients}
The effective channel coefficients and their variances are estimated by incorporating both the received pilot and data signals. In order to approximate the posterior distribution of $h_{mn}$ in the $i$-th iteration of the joint estimator, we first obtain the belief of variable node $h_{mn}$ as it only differs from the posterior distribution of $h_{mn}$ by a normalizing constant, which can be derived based on the BP algorithm as follows:
\begin{align}
    B_{h_{mn}}^{(j)}(i) = I_{f_{h_{mn} \rightarrow h_{mn}}}^{(j)}(i) \prod_{t \in \mathbf{T}_{p}} I_{f_{y_{mt} \rightarrow h_{mn}}}^{(j)}(i) \prod_{t \in \mathbf{T}_{d}} I_{f_{y_{mt} \rightarrow h_{mn}}}^{(j)}(i),
\end{align}
\noindent where $I_{f_{h_{mn} \rightarrow h_{mn}}}^{(j)}(i)$ denotes the message from factor node $p(h_{mn}|u_{n})$ to variable node $h_{mn}$ that serves as the prior distribution of $h_{mn}$, and $I_{f_{y_{mt} \rightarrow h_{mn}}}^{(j)}(i)$ represents the message from factor node $p(y_{mt}|z_{mt})$ to variable node $h_{mn}$. Thus, the term $\prod_{t \in \mathbf{T}_{p}} I_{f_{y_{mt} \rightarrow h_{mn}}}^{(j)}(i)$ $\prod_{t \in \mathbf{T}_{d}} I_{f_{y_{mt} \rightarrow h_{mn}}}^{(j)}(i)$ can be interpreted as the likelihood function of $h_{mn}$ in the $i$-th iteration. Specifically, the term $\prod_{t \in \mathbf{T}_{p}} I_{f_{y_{mt} \rightarrow h_{mn}}}^{(j)}(i)$, which corresponds to the received pilot symbols, is approximated as a complex Gaussian PDF with mean $P_{p,mn}^{h(j)}(i)$ and variance $Q_{p,mn}^{h(j)}(i)$, and the term $\prod_{t \in \mathbf{T}_{d}} I_{f_{y_{mt} \rightarrow h_{mn}}}^{(j)}(i)$ that relates to the received data symbols, is approximated as another complex Gaussian PDF with mean $P_{d,mn}^{h(j)}(i)$ and variance $Q_{d,mn}^{h(j)}(i)$. Consequently, the term $\prod_{t \in \mathbf{T}_{p}} I_{f_{y_{mt} \rightarrow h_{mn}}}^{(j)}\!(i)\!\prod_{t \in \mathbf{T}_{d}} I_{f_{y_{mt} \rightarrow h_{mn}}}^{(j)}\!(i)$ is also approximated as a complex Gaussian PDF with mean $P_{mn}^{h(j)}(i)$ and variance $Q_{mn}^{h(j)}(i)$ given in Line 13 and 14 of Algorithm 2, respectively, which are derived in Appendix B. 

To derive $B_{h_{mn}}^{(j)}(i)$, we obtain $I_{f_{h_{mn} \rightarrow h_{mn}}}^{(j)}(i)$ as follows:
\begin{align}
I_{f_{h_{mn} \rightarrow h_{mn}}}^{(j)}(i) = \left(1-\rho_{mn}^{(j)}(i)\right)\delta(h_{mn})\!+\!\rho_{mn}^{(j)}(i)\mathcal{CN}(h_{mn};0,\beta_{n}),
\end{align}
\noindent where $\rho_{mn}^{(j)}(i)$ approximates the probability that $h_{mn}$ is non-zero, and it is defined as the \emph{sparsity level} of user $n$ at the $m$-th BS antenna. Detailed derivations of (16) are deferred to Appendix C. Note that estimates of the likelihood that each user is active or not in the considered transmission block, i.e., $\lambda_n^{\left(j\right)}$, are required for the calculations of $\{\rho_{mn}^{(j)}\left(i\right)\}$'s in Lines 15-17 of Algorithm 2. We propose to update $\{\lambda_{n}^{\left(j\right)}\}$'s in each turbo iteration for more accurate estimation of the BiG-AMP algorithm, as will be elaborated shortly. Therefore, the posterior distribution of $h_{mn}$ is approximated in the $i$-th iteration as follows:
\begin{align}
\begin{split}
r_{h_{mn}}^{(j)}(i)&= \frac{B_{h_{mn}}^{(j)}(i)}{\int B_{h_{mn}}^{(j)}(i) dh_{mn}}=\left(1-\tilde{\rho}_{mn}^{(j)}(i)\right)\delta(h_{mn})\\
&+\tilde{\rho}_{mn}^{(j)}(i)\mathcal{C}\mathcal{N}\left(h_{mn};\mu_{mn}^{(j)}(i), \tau_{mn}^{(j)}(i)\right),
\end{split}
\end{align}

\noindent where $\mu_{mn}^{(j)}(i)$ and $\tau_{mn}^{(j)}(i)$ are given in Line 18 of Algorithm 2, and $\tilde{\rho}_{mn}^{(j)}(i)$ presented in Line 19 is defined as the \emph{posterior sparsity level} of user $n$ at the $m$-th BS antenna. Based on the posterior distribution, the MMSE estimate of the effective channel coefficient and its variance are obtained in Line 20 and 21 of Algorithm 2, respectively.

\subsubsection{Estimate the soft data symbols}
Due to the symmetry of $x_{nt}$ and $h_{mn}$ in the bilinear inference problem of the joint estimator, we similarly obtain the conditional mean $P_{nt}^{x(j)}(i)$ and variance $Q_{nt}^{x(j)}(i)$ given $x_{nt}$ in Line 23 and 22 of Algorithm 2, respectively. Prior distributions of the transmitted data symbols can be estimated as follows:
\begin{align}
p(x_{nt})=I_{f_{x_{nt}}\rightarrow x_{nt}}^{(j)}(i) \approx \bar{\rho}_{n}^{(j-1)} \sum_{s \in \mathcal{X}}\eta_{nt,s}^{(j)}\delta(x_{nt}-s), t \in \mathbf{T}_{d}, 
\end{align}

\noindent where $\bar{\rho}_{n}^{(j-1)} \triangleq \frac{1}{M}\sum_{m\in\mathcal{M}} \tilde{\rho}_{mn}^{(j-1)}$, and $\eta_{nt,s}^{(j)}$ denotes the probability that $x_{nt}$ belongs to constellation point $s$. As will be introduced in the next subsection, $\{\eta_{nt,s}^{(j)}\}$'s are obtained from the channel decoder in the last turbo iteration. Thus, the approximated posterior distributions of the transmitted data symbols in the $i$-th iteration of the joint estimator can be expressed as follows:
\begin{align}
r_{x_{nt}}^{(j)}(i) = \bar{\rho}_{n}^{(j-1)} \sum_{s \in \mathcal{X}}\tilde{\eta}_{nt,s}^{(j)}(i)\delta(x_{nt}-s), t \in \mathbf{T}_{d}, 
\end{align}

\noindent where $\tilde{\eta}_{nt,s}^{(j)}(i)$ denotes the posterior probability that $x_{nt}$ belongs to constellation point $s$ as derived using the Bayes' rule in Line 24 of Algorithm 2. The soft data symbols and the corresponding posterior variances are estimated via Line 25 and 26, respectively.

Once the while loop of Algorithm 2 is terminated, $\bar{\rho}_{n}^{(j)}$, $\hat{h}_{mn}^{(j)}$ and $V_{mn}^{h(j)}$ are updated for the next turbo iteration. Accordingly, we update $\{\lambda_{n}^{(j)}\}$'s in Line 28 using the average sparsity level $\bar{\rho}_{n}^{(j)}$, i.e., $p(u_{n} = 1) \triangleq \lambda_{n}^{(j+1)}=\kappa \bar{\rho}_{n}^{(j)}+(1-\kappa)\lambda_{n}^{(j)}$, $n\in \mathcal{N}$, where $\kappa \in [0,1]$ is the \emph{learning rate}. This is inspired by the idea of \emph{exploration and exploitation} from reinforcement learning \cite{rssu2018}, which avoids using the average sparsity levels exclusively to eliminate the potential estimation errors caused by inaccurate prior information of the BiG-AMP algorithm. The set of active users is determined as $\hat{\Xi}^{(j)} \triangleq \{n\in\mathcal{N}|\bar{\rho}_{n}^{(j)} \geq\theta\}$, where $\theta$ is an empirical threshold \cite{mke2020}.

\subsubsection{Derive the extrinsic information of the joint estimator}
With the estimated set of active users $\hat{\Xi}^{(j)}$ and soft data symbols $\{\hat{x}_{nt}\}$'s, extrinsic information of the joint estimator is derived as input of the channel decoder, which aims at minimizing the data decoding error by eliminating some redundancy from the prior information of the coded bits. In particular, the posterior probability of a transmitted data symbols is translated to the posterior probabilities of the corresponding coded bits as follows:
\begin{align}
p\left(c_{nj_{c}}^{(j)}=b|\mathbf{Y}\right)=\sum \nolimits_{s \in \mathcal{X}_{\hat{j}_{c}}^{b}}\tilde{\eta}_{nt,s}^{(j)}, n \in \hat{\Xi}^{(j)},
\end{align}
\noindent where $\hat{j}_{c}\triangleq \mod({j}_c,\log_{2}|\mathcal{X}|)$, $t=L+1+\left\lfloor \frac{j_{c}}{\log_{2}|\mathcal{X}|} \right\rfloor$, and $\mathcal{X}_{l}^{b}$ represents the set of constellation points with the $l$-th position ($l=0,\cdots, \log_{2}|\mathcal{X}|-1$) of the corresponding bit sequence as $b$. For example, suppose the bit sequences “00”, “01”, “10” and “11” are modulated to constellation points $s_{0}$, $s_{1}$, $s_{2}$, and $s_{3}$ respectively in quadrature phase shift keying (QPSK), we have $\mathcal{X}_{0}^{0} = \{s_{0},s_{1}\}$, $\mathcal{X}_{0}^{1} = \{s_{2},s_{3}\}$, $\mathcal{X}_{1}^{0} = \{s_{0},s_{2}\}$, and $\mathcal{X}_{1}^{1} = \{s_{1},s_{3}\}$. Other modulation schemes such as the 16-Quadrature Amplitude Modulation (QAM) can be applied similarly. The posterior probabilities of the coded bits are used to derive the posterior LLRs as follows:
\begin{align}
L_{E}^{p}\left(c_{nj_{c}}^{(j)}\right)\triangleq \ln \left( \frac{p(c_{nj_{c}}^{(j)}=0|\mathbf{Y})}{p(c_{nj_{c}}^{(j)}=1|\mathbf{Y})}\right), n \in \hat{\Xi}^{(j)},
\end{align}

\noindent which are converted to the extrinsic information as defined below \cite{bvu2001}:
\begin{align}
L_{E}^{e}\left(c_{nj_{c}}^{(j)}\right)\triangleq L_{E}^{p}\left(c_{nj_{c}}^{(j)}\right)-L_{E}^{a}\left(c_{nj_{c}}^{(j)}\right), n \in \hat{\Xi}^{(j)},
\end{align}

\noindent where $L_{E}^{a}\left(c_{nj_{c}}^{(j)}\right)\triangleq \ln \left( \frac{p(c_{nj_{c}}^{(j)}=0)}{p(c_{nj_{c}}^{(j)}=1)}\right)$ is the prior information obtained from the channel decoder in the last turbo iteration.
\subsection{The Channel Decoder}
The channel decoder determines the most probable code block for each user that is determined as active by the joint estimator, which can be formulated as the following maximum \emph{a posteriori} probability (MAP) estimation problem:
\begin{align}
\bm{\hat{c}}_{n}=\arg \max_{\bm{c}_{n} \in \{0,1\}^{N_{c}}} p\left(\bm{c}_{n} \mid \{L_{E}^{e}(c_{nj_{c}}^{(j)})\}\right), n \in \hat{\Xi}^{(j)}.
\end{align}

Known for its effectiveness in calculating marginal distributions, the BP algorithm has a long history of applications for channel decoder designs \cite{bvu2001}. In this paper, we adopt a BP-based channel decoder to solve (23), which should be able to accept the extrinsic information of the coded bits derived from the joint estimator as input, and calculate the posterior LLRs of the coded bits $L_{D}^{p}(c_{nj_{c}}^{(j)})$, $n \in \hat{\Xi}^{(j)}$ as the soft decoding results. We emphasize that this is a mild requirement that can be satisfied by a variety of off-the-shelf BP-based channel decoders, e.g., the decoders developed in \cite{rgga1962}, \cite{hoc2004} and \cite{cbe1996}, \cite{jha1996} for low-density parity-check (LDPC) code and turbo code, respectively. 

Similar to the joint estimator, extrinsic information of the channel decoder is derived as follows:
\begin{align}
L_{D}^{e}\left(c_{nj_{c}}^{(j)}\right)\triangleq L_{D}^{p}\left(c_{nj_{c}}^{(j)}\right)-L_{D}^{a}\left(c_{nj_{c}}^{(j)}\right), n \in \hat{\Xi}^{(j)},
\end{align}
\noindent which is adopted as prior information $L_{E}^{a}(c_{nj_{c}}^{(j+1)})$, $n \in \hat{\Xi}^{(j)}$ for the use of the joint estimator in the next turbo iteration. Therefore, the prior distribution of a coded bit is given as
\begin{align}
p\left(c_{nj_{c}}^{(j)}\right)=\left\{
\begin{aligned}
\frac{1}{1+\text{exp}\left(L_{D}^{e}\left(c_{nj_{c}}^{(j)}\right)\right)}, c_{nj_{c}}^{(j)} = 1,\\
\frac{\text{exp}\left(L_{D}^{e}\left(c_{nj_{c}}^{(j)}\right)\right)}{1+\text{exp}\left(L_{D}^{e}\left(c_{nj_{c}}^{(j)}\right)\right)}, c_{nj_{c}}^{(j)} = 0,
\end{aligned}
\right. \ n \in \hat{\Xi}^{(j)},
\end{align} 
\noindent and prior distributions of the transmitted data symbols can be estimated according to the following expression:
\begin{align}
{\eta}_{nt,s}^{(j+1)}=\prod \nolimits_{j_{c}=v_{1}}\nolimits^{v_{2}}p\left(c_{nj_{c}}^{(j)}\right), t \in \mathbf{T}_{d}, n\in \hat{\Xi}^{(j)}.
\end{align}

\noindent where $v_{1}\triangleq (t-L-1)\text{log}_{2}|\mathcal{X}|$, $v_{2}\triangleq (t-L)\text{log}_{2}|\mathcal{X}|-1$, and $\mu\left([c_{nv_{1}},\cdots,c_{nv_{2}}] \right) = s$. Note that for the users that are determined as inactive, we reuse the prior information of the transmitted data symbols from the last turbo iteration by setting $L_{E}^{a}\left(c_{nj_{c}}^{(j+1)}\right)=L_{E}^{a}\left(c_{nj_{c}}^{(j)}\right)$ and ${\eta}_{nt,s}^{(j+1)}={\eta}_{nt,s}^{(j)}$, $n \in  \mathcal{N}\setminus \hat{\Xi}^{(j)}$.

The values of $\{L_{D}^{p}\left(c_{nj_{c}}\right)\}$'s are also utilized to obtain the code block $\hat{\bm{d}}_{n}$, $n \in \hat{\Xi}$ after the last turbo iteration by performing hard decision as follows:
\begin{align}
\hat{d}_{nj_{c}}=\left\{
\begin{aligned}
0,  L_{D}^{p}\left(c_{nj_{c}}\right)\geq 0, \\
1,  L_{D}^{p}\left(c_{nj_{c}}\right)<0.
\end{aligned}
\right.
\end{align}
\noindent CRC is then performed for $\hat{\bm{d}}_{n}, n\in\hat{\Xi}$ and the CRC bits are detached to obtain the payload bits $\hat{\bm{b}}_{n}$ for the users that pass parity check, which is denoted as $\hat{\Xi}_{c}$ in Algorithm 1.

\section{A Low-complexity Side Information-aided Receiver}
Assisted by prior information of the transmitted data symbols, the proposed turbo receiver effectively exploits the common sparsity pattern in the received pilot and data signal via BiG-AMP. Nevertheless, such a design incurs significant computation overhead even with a reasonable size of the payload data, since the joint estimation of effective channel coefficients and soft data symbols is performed iteratively by incorporating all the received symbols in each turbo iteration. Besides, in order to estimate the prior information, the channel decoder needs to be executed in each turbo iteration for all users in $\hat{\Xi}^{\left(j\right)}$ (See Line 7 of Algorithm 1), which brings additional computation overhead. These observations necessitate low-complexity receivers for massive RA that can leverage both the common sparsity pattern and channel decoding results more efficiently. In this section, we develop a low-complexity side information (SI)-aided receiver without relying on BiG-AMP.

\subsection{Overview of the SI-aided Receiver}
\begin{figure}[t]
\centering
\includegraphics[width=3.4in]{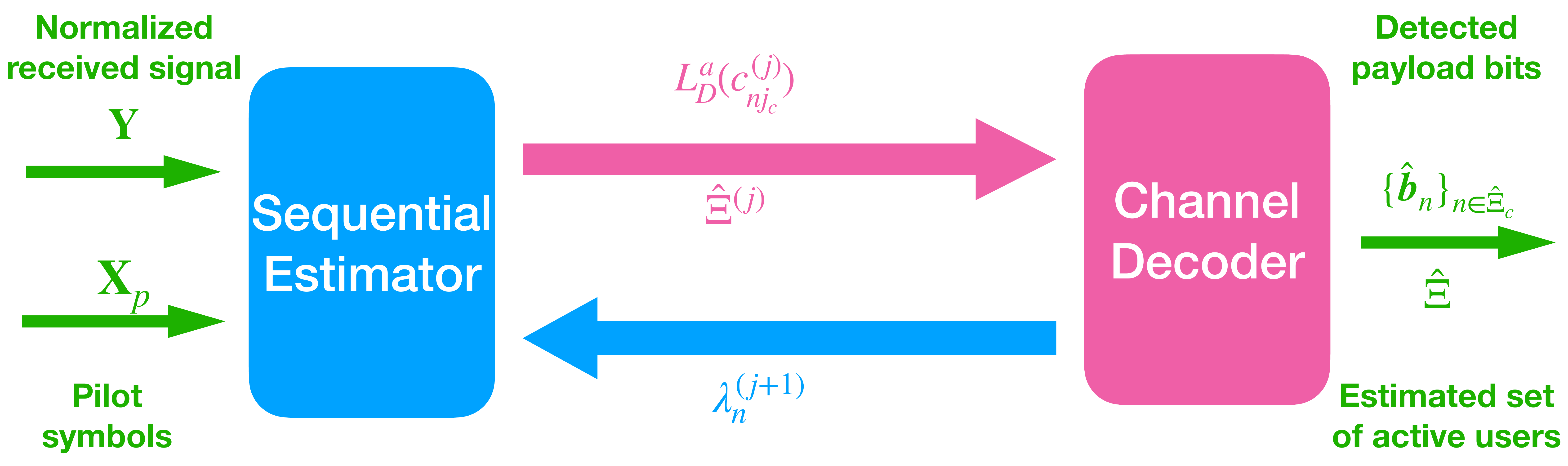}
\caption{The proposed SI-aided receiver for massive RA.}
\label{lowmodel}
\end{figure}

The SI-aided receiver iterates between a sequential estimator and a channel decoder as shown in Fig. \ref{lowmodel}. Unlike the turbo receiver developed in Section \uppercase\expandafter{\romannumeral3}, it estimates the effective channel coefficients and soft data symbols sequentially in each iteration to reduce the computation overhead. Specifically, the sequential estimator cascades the AMP algorithm \cite{mke2020} for JADCE, and an MMSE-based soft demodulator to compute the prior LLRs of the coded bits. A BP-based channel decoder is adopted to obtain the posterior LLRs of the coded bits similar as the turbo receiver, while hard decision is required for parity check in each iteration. By updating the SI, i.e., the estimates on whether a user is active, in each iteration jointly based on the average sparsity levels, the posterior LLRs of the coded bits, and the parity check results, the receiver progresses with more precise prior knowledge for the sequential estimator so that more accurate JADCE can be achieved \cite{ama2019}. The workflow of the SI-aided receiver is summarized in Algorithm 3. We introduce details of the sequential estimator and the channel decoder in Section \uppercase\expandafter{\romannumeral4}-B, and elaborate the design of the SI in Section \uppercase\expandafter{\romannumeral4}-C.

\begin{algorithm}[htbp]
\caption{The Proposed SI-aided Receiver for Massive RA}
{\bf Input:}
The normalized received signal $\mathbf{Y}$, pilot symbols $\mathbf{X}_{p}$, maximum number of iterations $Q_{3}$, and accuracy tolerance $\epsilon_{3}$.\\
{\bf Output:} 
The estimated set of active users $\hat{\Xi}$, the set of users that pass CRC $\hat{\Xi}_{c}$ and their detected payload bits $\hat{\bm{b}}_{n}$.\\
{\bf Initialize:}
$\!j\!\leftarrow\!0$, $\lambda_{n}^{(1)}\!\leftarrow\!\frac{K}{N}$, $n\!\in\!\mathcal{N}$, $\hat{x}_{nt}^{(0)} \leftarrow 0$, $t \in \mathbf{T}_{d}$, $\hat{\Xi}_{c} \leftarrow \emptyset$.
\begin{algorithmic}[1]
\WHILE{{$j < Q_{3}$} \text{and} {$\frac{\Sigma_{n,t}|\hat{x}_{n t}^{(j)}-\hat{x}_{n t}^{(j-1)}|^{2}}{\Sigma_{n,t}|\hat{x}_{n t}^{(j-1)}|^{2}} > \epsilon_{3}$}}
\STATE $j \leftarrow j+1$
\Statex \quad //\textit{The Sequential Estimator}//
\STATE Execute the AMP algorithm \cite{mke2020} with the SI $\{\lambda_{n}^{(j)}\}$'s

\Statex \quad as the prior knowledge of the user activity to estimate 
\Statex \quad the effective channel coefficients $\{\hat{h}_{mn}^{(j)}\}$'s and set of 

\Statex \quad active users $\hat{\Xi}^{(j)}$.

\STATE Estimate the transmitted data symbols $\hat{x}_{nt}^{(j)}$, $t \in \mathbf{T}_{d}$ via

\Statex \quad an MMSE equalizer, i.e., $\hat{\mathbf{X}}_{d, a}^{(j)}=\Big((\hat{\mathbf{H}}_{a}^{(j)})^{\mathrm{H}} \hat{\mathbf{H}}_{a}^{(j)}+$

\Statex \quad $(\sigma^{2}/\gamma)\mathbf{I}\Big)^{-1} (\hat{\mathbf{H}}_{a}^{(j)})^{\mathrm{H}} \mathbf{Y}_{d}$, where $\mathbf{H}_{a}^{(j)} \triangleq \left[\left\{\hat{\mathbf{h}}_{k}^{(j)}\right\}_{k \in \hat{\Xi}^{(j)}}\right]$

\Statex \quad stacks the effective channel coefficients of all 

\Statex \quad the estimated active users, and $\hat{x}_{nt}^{(j)}$ is the entry of $\hat{\mathbf{X}}_{d, a}^{(j)}$.

\STATE Compute the prior LLRs of the coded bits as $L_{D}^{a}\left(c_{nj_{c}}^{(j)}\right)$,

\Statex \quad $n \in \hat{\Xi}^{(j)}$ via soft demodulation according to (28).

\Statex \quad //\textit{The Channel Decoder}//
\STATE Perform soft data decoding via a BP-based channel

\Statex \quad decoder to obtain the posterior LLRs of the coded bits 

\Statex \quad $L_{D}^{p}\left(c_{nj_{c}}^{(j)}\right)$, $n \in \hat{\Xi}^{(j)} \setminus \hat{\Xi}_{c}$.
\STATE Perform hard decision, determine the set of users in
\Statex \quad $\hat{\Xi}^{(j)}$ that pass $\hat{\Xi}_{c}^{(j)}$ and obtain their payload bits $\{\hat{\bm{b}}_{n}\}$'s.

\STATE $\hat{\Xi}_{c} \leftarrow \hat{\Xi}_{c} \bigcup \hat{\Xi}_{c}^{(j)}$
\STATE Update the SI $\{\lambda_{n}^{(j+1)}\}$'s according to (29).
\ENDWHILE
\end{algorithmic}
\end{algorithm}

\subsection{The Sequential Estimator and the Channel Decoder}
Based on the normalized received pilot signal $\mathbf{Y}_{p}$, the sequential estimator adopts the AMP algorithm \cite{mke2020} to estimate the effective channel coefficients $\{\hat{h}_{mn}^{(j)}\}$'s and the set of active users $\hat{\Xi}^{(j)}$. We also derive the sparsity levels $\{\rho_{mn}^{(j)}\}$'s from the AMP algorithm following similar steps in Lines 15-17 of Algorithm 2. Soft data symbol detection is then performed using an MMSE-based soft demodulator based on the results of the sequential estimator. In particular, signal distortion caused by wireless fading is first removed from the normalized received data signal $\mathbf{Y}_{d}$ to estimate the transmitted data symbols $\hat{x}_{nt}$, $n \in \hat{\Xi}^{(j)}$, $t \in \mathbf{T}_{d}$ via an MMSE equalizer. The prior LLRs of the coded bits are obtained via soft demodulation as follows:
\begin{align}
\begin{split}
L_{D}^{a}\left(c_{nj_{c}}^{(j)}\right)&\triangleq \ln \left(\frac{p\left(c_{nj_{c}}^{(j)}=0|\hat{x}_{nt}^{(j)}\right)}{p\left(c_{nj_{c}}^{(j)}=1|\hat{x}_{nt}^{(j)}\right)}\right)\\
&= \ln \left( \frac{\sum_{s \in \mathcal{X}_{\hat{j}_{c}}^{0}}\exp\left(-\gamma||\hat{x}_{nt}^{(j)}-s||_{2}^{2}/\sigma^{2}\right)}{\sum_{s \in \mathcal{X}_{\hat{j}_{c}}^{1}}\exp\left(-\gamma||\hat{x}_{nt}^{(j)}-s||_{2}^{2}/\sigma^{2}\right)}\right)
, n \in \hat{\Xi}^{(j)},
\end{split}
\end{align}

\noindent where $\hat{j}_{c}\triangleq \mod({j}_c,\log_{2}|\mathcal{X}|)$ and $t=L+1+\left\lfloor \frac{j_{c}}{\log_{2}|\mathcal{X}|} \right\rfloor$.

With the knowledge of $\{L_{D}^{a}\left(c_{nj_{c}}^{(j)}\right)\}$'s, the BP-based channel decoder calculates the posterior LLRs of the coded bits $L_{D}^{P}\left(c_{nj_{c}}^{(j)}\right)$, $n \in \hat{\Xi}^{(j)} \setminus \hat{\Xi}_{c}$ and decides the code blocks according to (27). CRC is performed for all users in $\hat{\Xi}^{(j)} \setminus \hat{\Xi}_{c}$ to obtain their payload bits $\{\hat{\bm{b}}_{n}\}$'s. Note that channel decoding is not performed for the users that have already passed the parity check, which differs the turbo receiver and helps to save the computations. We would like to point out that the SI-aided receiver can be extended by incorporating the idea of \emph{successful interference cancellation}, i.e., subtracting the user data that have passed CRC from the received signal. However, the performance improvement is not guaranteed due to the potentially large channel estimation error. A thorough investigation on such an extension will be left for future works.

\subsection{The Side Information}
The AMP algorithm is a key component of the sequential estimator, which determines the average sparsity levels and effective channel coefficients based on the framework of Bayesian estimation \cite{smkay1993}. As a consequence, prior knowledge of the user activity, i.e., the SI for the sequential estimator $\{\lambda_{n}\}$'s, also has significant impacts on the estimation accuracy, similar with the case of the BiG-AMP algorithm. In order to obtain more precise estimates through multiple iterations of Algorithm 3, we propose to update the SI by jointly utilizing the results of the sequential estimator and channel decoder, according to three different cases depending on the estimated set of active users and their parity check results via the following update rule:
\begin{align}
    \lambda_{n}^{(j+1)}=\left\{
\begin{aligned}
& 1, \ n \in \hat{\Xi}_{c}, \\
& \kappa_{1}\bar{\rho}_{n}^{(j)}+\frac{1-\kappa_{1}}{N_{c}}\sum_{j_{c}}\frac{\left|L_{D}^{P}(c_{nj_{c}}^{(j)})\right|}{1+\left|L_{D}^{P}(c_{nj_{c}}^{(j)})\right|}, \ n \in \hat{\Xi}^{(j)} \setminus \hat{\Xi}_{c}, \\
& \kappa_{2}\bar{\rho}_{n}^{(j)}+(1-\kappa_{2})\lambda_{n}^{(j)}, \ n \in  \mathcal{N}\setminus (\hat{\Xi}^{(j)} \cup \hat{\Xi}_{c}).
\end{aligned}
\right.
\end{align}

\noindent In particular, in the first case of (29), we set $\lambda_{n}^{(j+1)}=1$, $n \in \hat{\Xi}_{c}$ since the users that have passed the parity check in the current or previous iterations can be safely determined as active. In the second case, we handle the users that are estimated as active but fail to pass the parity check in the current iteration, i.e., $n \in \hat{\Xi}^{(j)} \setminus \hat{\Xi}_{c}$. For this set of users, the average sparsity levels and the posterior LLRs of the coded bits are jointly utilized to update the SI since both of them are informative on users' activity. Specifically, the term $\frac{1}{N_{c}}\sum_{j_{c}}\frac{\left|L_{D}^{P}(c_{nj_{c}}^{(j)})\right|}{1+\left|L_{D}^{P}(c_{nj_{c}}^{(j)})\right|} \in [0,1)$ indicates the decoding reliability of user $n$ as its complement, i.e., $\frac{1}{N_{c}}\sum_{j_{c}}\frac{1}{1+\left|L_{D}^{P}(c_{nj_{c}}^{(j)})\right|}$, provides an accurate estimate of the bit error rate \cite{bgo2018}. Besides, parameter $\kappa_{1} \in [0,1]$ is an empirical weighting factor balancing the contributions of the channel estimation and data decoding results. In the third case, we update the SI for the users that neither pass the CRC in any iteration nor being determined as inactive in the current iteration using the average sparsity levels, using similar methodology as that for the turbo receiver in Section \uppercase\expandafter{\romannumeral3}, where $\kappa_{2} \in [0,1]$ denotes the learning rate.

\begin{figure}[htpb]
\centering
\includegraphics[width=3in]{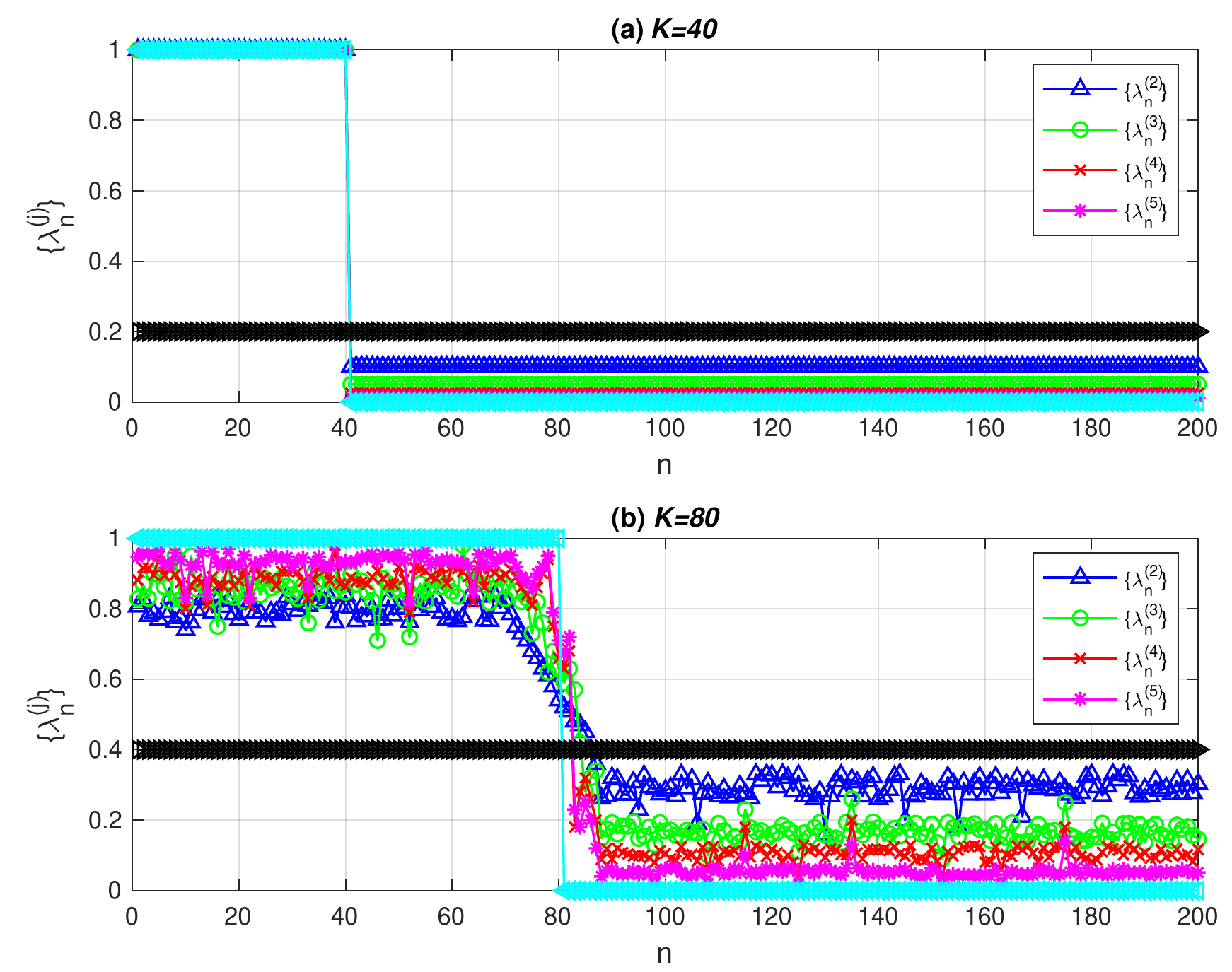}
\caption{Illustrations on the SI update rule in (29). We set $N=200$ and $\kappa_{1}=\kappa_{2}=0.5$. The first $K$ users are active in the transmission block.}
\label{evolve}
\end{figure}

To demonstrate the rationality of the SI update rule in (29), we provide numerical examples on the evolutions of $\{\lambda_{n}^{(j)}\}$'s, considering two scenarios with $K=40$ and $80$ in Fig. \ref{evolve}(a) and (b), respectively. From these figures, we see that as the iteration of Algorithm 3 proceeds, the SI evolves from the initial values, which is set to be $\lambda_{n}^{(0)}=\frac{K}{N}$, $n\in\mathcal{N}$, to the perfect estimates, i.e., $\lambda_{n}^{(0)}=1$, $n=1, \cdots, K$ and $\lambda_{n}^{(0)}=0$, $n=K+1, \cdots, N$. This validates the effectiveness of the proposed SI update rule. Besides, we observe that a larger number of active users leads to a slower convergence rate and higher estimation variance, implying the need of more iterations in Algorithm 3.

\subsection{Computational Complexity Analysis}
The computational complexity of the two proposed receivers is summarized in TABLE \ref{complexity}, where the number of complex-valued multiplications is adopted as the metric, and the complexity of a real-valued multiplication is assumed to be one quarter of a complex-valued multiplication. We use $O_{d}$ to denote the complexity of the decoder. Since the overall computational complexity of the two proposed receivers is determined by the iteration numbers, i.e., $Q_{1}$ and $Q_{3}$ for the turbo and SI-aided receiver respectively, we only focus on the complexity of one iteration in the following discussions, which is contributed by the estimator and channel decoder.
\begin{table}[ht]
\caption{Complexity of the proposed receivers in one iteration}
\centering
\scalebox{1}{
    \begin{tabular}{c|c}
    \hline
    Receiver & Number of complex multiplications\\
    \hline
    Turbo & \makecell{$(\frac{9}{4}MNL+\frac{13}{4}MNL_{d}+\frac{15}{4}MN$\\ $+\frac{3}{2}M(L+L_{d})+\frac{3}{4}NL_{d}|\mathcal{X}|)T_{j}+K_{1}'O_{d}$}\\
    \hline
    SI-aided & \makecell{$(4MNL+\frac{7}{4}MN+\frac{19}{2}NL)T_{s}+\frac{5}{2}K_{2}^{2}M$\\$+\frac{3}{4}MK_{2}L_{d}+\frac{5}{4}K_{2}+\frac{1}{2}K_{2}L_{d}|\mathcal{X}|+K_{2}'O_{d}$}\\
    \hline
    \end{tabular}}
\label{complexity}
\end{table}

As summarized in TABLE \ref{complexity}, the complexity of the channel decoder in each turbo iteration is given as $K_{1}'O_{d}$ for the turbo receiver ($K_{2}'O_{d}$ for the SI-aided receiver), where $K_{1}^{\prime}$ ($K_{2}^{\prime}$) is the number of users that need to be decoded in this iteration. The other terms shown in the table correspond to the complexity of the joint (sequential) estimator. Compared with the joint estimator that requires $(\frac{9}{4}MNL+\frac{13}{4}MNL_{d}+\frac{15}{4}MN+\frac{3}{2}M(L+$ $L_{d})+\frac{3}{4}NL_{d}|\mathcal{X}|)T_{j}$ complex-valued multiplications for the BiG-AMP algorithm in each turbo iteration, the sequential estimator first performs the AMP algorithm, followed by an one-shot data detection procedure, which respectively require $(4MNL+\frac{7}{4}MN+\frac{19}{2}NL)T_{s}$ and $\frac{5}{2}K_{2}^{2}M+\frac{3}{4}MK_{2}L_{d}+\frac{5}{4}K_{2}+\frac{1}{2}K_{2}L_{d}|\mathcal{X}|$ complex-valued multiplications. Please be noted that $K_2$ denotes the number of users that are detected as active, and $T_j$ and $T_s$ stand for the actual iteration numbers of the BiG-AMP and AMP algorithm, respectively.

We use the simulation setting as will be detailed in Section V to give a more intuitive idea on the computational complexity of the two proposed receivers by assuming $K=20$. The values of $T_{j}$ and $T_{s}$, given respectively by 41 and 52, are obtained by averaging the results over $100$ independent channel realizations. The numbers of complex-valued multiplications of the turbo and SI-aided receiver in one iteration (excluding those of the channel decoders) are given by $3.2\times 10^8$ and $1.39 \times 10^8$, respectively. In addition, since hard decision is made in each iteration of the SI-aided receiver, $K_{2}'$ is typically smaller than $K_{1}^{\prime}$, especially in later iterations. These two factors jointly imply that the SI-aided receiver has a much lower complexity compared with the turbo receiver. In the next section, we will compare the computational complexity of different receivers numerically using the measured execution time in simulations.

\section{Simulation Results}
\subsection{Simulation Setting and Baseline Schemes}
A single-cell uplink cellular network is simulated, where 200 users are randomly distributed in a circle with a radius of 500 m centered at the BS equipped with 64 antennas. The path loss of user $n$ is calculated as $\beta_{n}=-128.1-36.7\text{log}_{10}(r_{n})$ (dB), where $r_{n}$ (km) is the distance to the BS. The system bandwidth is 1 MHz, and the user transmit power is 23 dBm. Without otherwise specified, QPSK is employed as the modulation scheme and LDPC code is used for channel coding. Besides, we select CRC-8 to show the effectiveness of the proposed receivers, which is one of the CRC options for the physical uplink control channel (PUCCH) in 3GPP standards \cite{3gpp2018}.

In order to achieve more stable convergence behavior, a damping factor $\omega \in (0,1]$ \cite{jtpa2014} is applied to moderate the updates of $M_{mt}^{p}$, $V_{mt}^{p}$, $\hat{h}_{mn}$, and $\hat{x}_{nt}$. For instance, the damped version of the estimated soft data symbol can be expressed as $\bar{x}_{nt}(i)=\omega\hat{x}_{nt}(i)+(1-\omega)\bar{x}_{nt}(i-1), t \in \mathbf{T}_{d}$. In particular, the mean and variance of $z_{mt}$ in Line 3 and 4 of Algorithm 2 are replaced with the damped versions $\bar{M}_{mt}^{p}$ and $\bar{V}_{mt}^{p}$, respectively, while the damped versions of $\hat{h}_{mn}$ and $\hat{x}_{nt}$ are used in Lines 10-12 and Lines 22-23 of Algorithm 2. The simulation results are averaged over $10^5$ independent channel realizations, and other critical simulation parameters are summarized in TABLE \ref{table1}.
\begin{table}[ht]
\caption{Simulation parameters}
\centering
    \begin{tabular}{c|c|c|c}
    \hline
     Parameters & Values & Parameters & Values \\
    \hline
    $M$ & 64 & $T$ & 200 \\
    \hline
    $L$ & 50 & $L_{d}$ & 150\\
    \hline
    $N_{b}$ & 142 & $N_{d}$ & 150\\
    \hline
    $N_{c}$ & 300 & CRC type & CRC-8 \\
    \hline
    $\omega$ & 0.6 & $N$ & 200\\
    \hline
    $\theta$ & 0.4 & $Q_{1}$, $Q_{3}$ & 6\\
    \hline
    $Q_{2}$ & 100 & $\epsilon_{1}$, $\epsilon_{2}$, $\epsilon_{3}$ & $10^{-5}$\\
    \hline
    $\kappa$, $\kappa_{2}$ & 0.5 & $\kappa_{1}$ & 0.5\\
    \hline
    Noise power density & -169 dBm/Hz & Code rate & $1/2$\\
    \hline
    \end{tabular}
\label{table1}
\end{table}

We adopt two baseline schemes and a performance upper bound for comparisons:
\begin{itemize}
    \item \textbf{Separate design \cite{mke2020}:} This scheme first performs JADCE via the AMP algorithm, after which, data symbols are detected using an MMSE equalizer. The detected soft data symbols are then converted to prior LLRs of the coded bits for data decoding via soft demodulation using (28). This can be viewed as an instance of the SI-aided receiver by setting $Q_{3}=1$. 
\end{itemize}

\begin{itemize}
    \item \textbf{Data-assisted design with BiG-AMP \cite{qzou2020}:} This scheme exploits the common sparsity pattern using the BiG-AMP algorithm for joint activity detection, channel estimation, and soft data symbol detection. The detected soft data symbols are converted to prior LLRs of the coded bits for data decoding using (28). This is a special case of the turbo receiver when $Q_{1}=1$. 
\end{itemize}

\begin{itemize}
    \item \textbf{Turbo receiver with known user activity:} This scheme assumes perfect knowledge of the user activity and consequently, channel estimation and data decoding are performed via the proposed turbo receiver by setting $\lambda_{n}^{(j)}=1$, $n \in \Xi$, and $\lambda_{n}^{(j)}=0$, $n\in \mathcal{N}\setminus \Xi$. This scheme serves as a performance upper bound.
\end{itemize}

\noindent Note that all the simulated schemes adopt the same BP-based LDPC decoder \cite{rgga1962} for fair comparisons.

\subsection{Results}
\begin{figure}[htbp]
\centering
\includegraphics[width=3in]{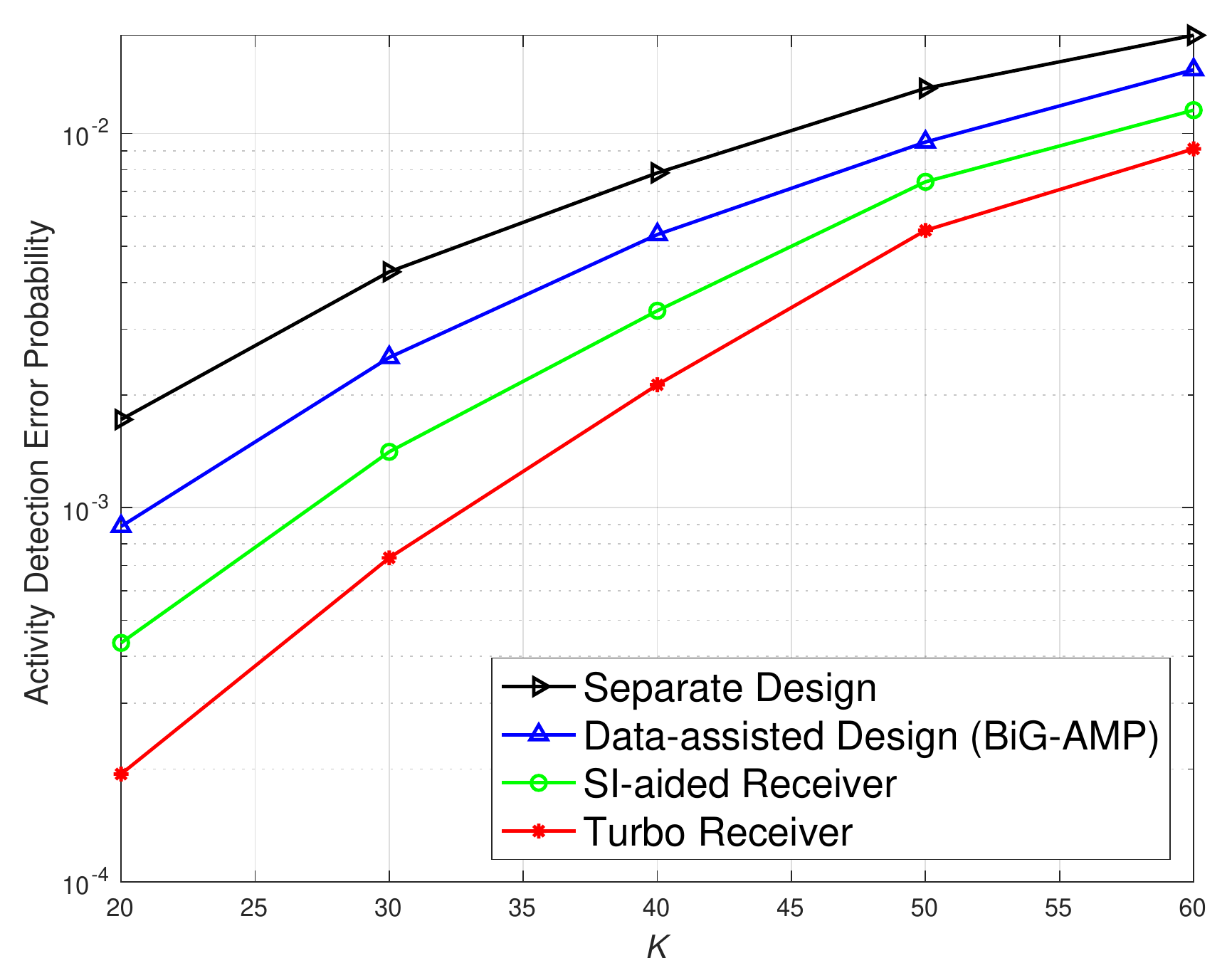}
\caption{Activity detection error probability vs. the number of active users.}
\label{AEP1}
\end{figure}

\begin{figure}[htbp]
\centering
\includegraphics[width=3in]{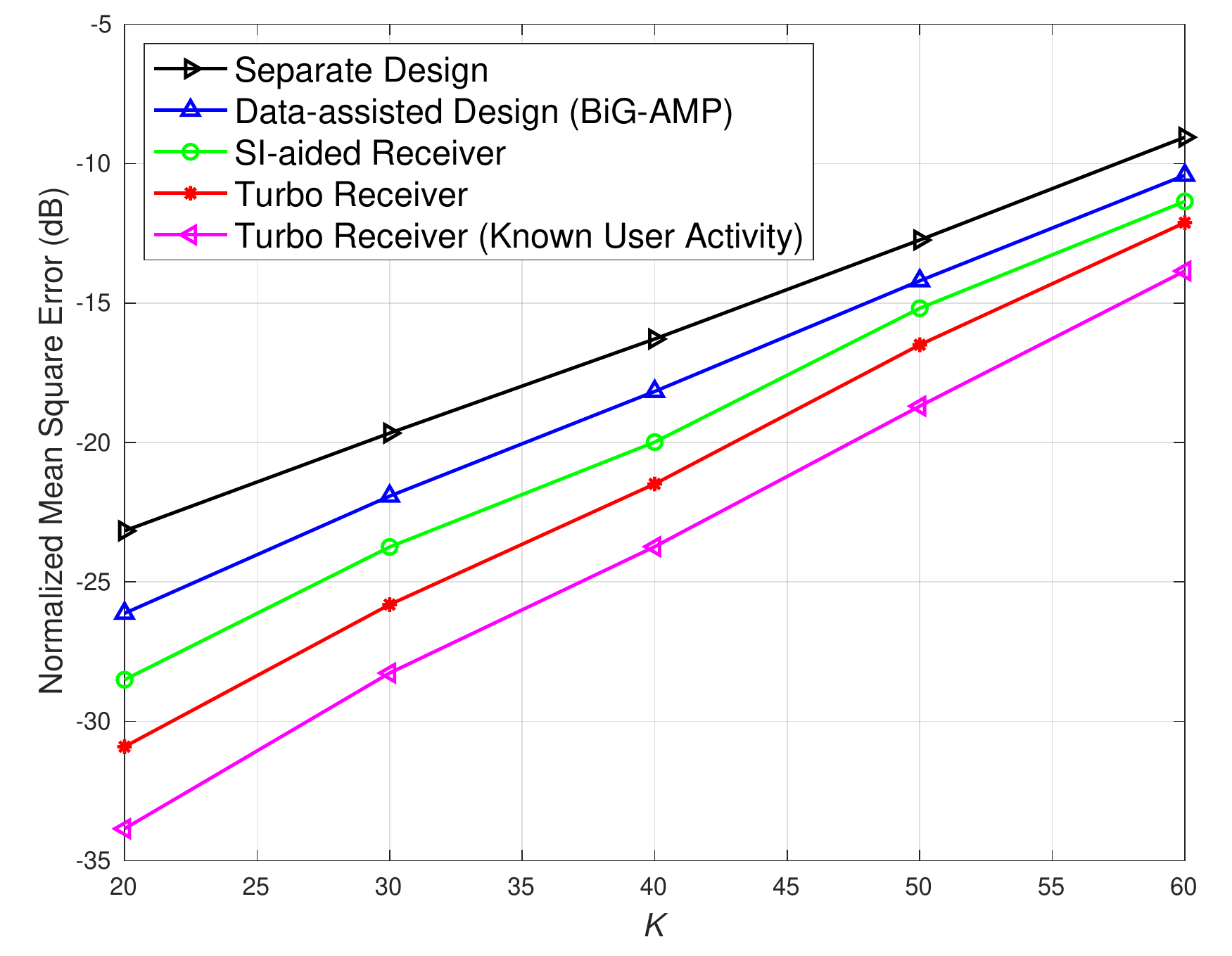}
\caption{NMSE of channel estimation vs. the number of active users.}
\label{NMSE1}
\end{figure}

We first evaluate the activity detection error probability (including the missed detection and false alarm probability) and the normalized mean square error (NMSE) of channel estimation in Fig. \ref{AEP1} and Fig. \ref{NMSE1}, respectively. It is observed that a large number of active users degrade both the activity detection and channel estimation performance due to the limited radio resource reserved for pilot transmission. Compared with the separate design, the data-assisted design achieves much lower activity detection and channel estimation errors, validating the benefits of incorporating the received data symbols. It is also seen that the proposed turbo receiver significantly outperforms the data-assisted design as the soft channel decoding results are further utilized to refine the prior distributions of the transmitted data symbols through multiple turbo iterations. Besides, despite with some performance loss compared with the turbo receiver, the low-cost SI-aided receiver secures noticeable performance improvement compared with the data-assisted design, which can be credited to the use of the customized SI as prior knowledge of the user activity.

\begin{figure}[htpb]
\centering
\includegraphics[width=3in]{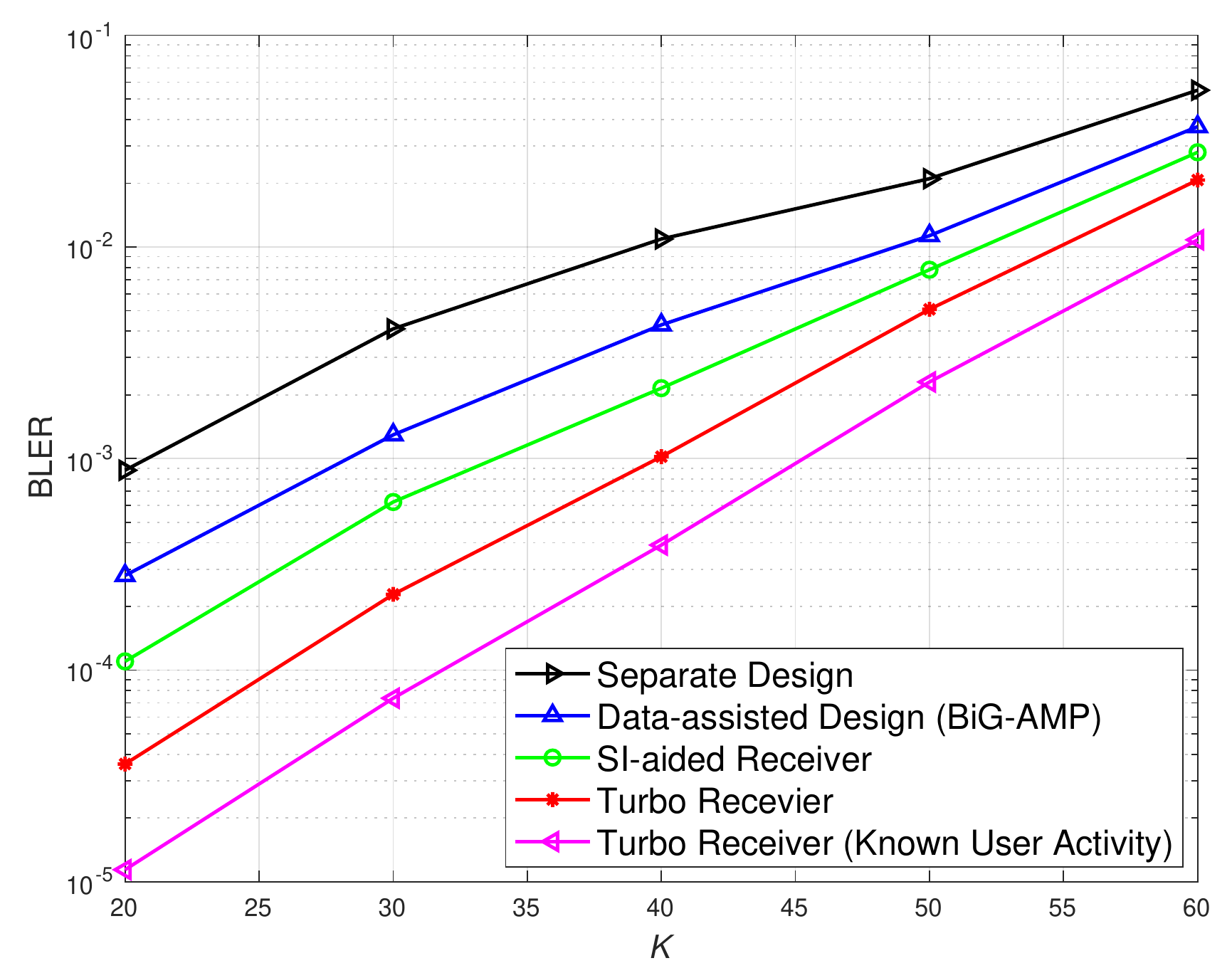}
\caption{BLER vs. the number of active users.}
\label{BLER1}
\end{figure}

Fig. \ref{BLER1} shows the BLER of all the simulated schemes versus the number of active users. Similar to activity detection and channel estimation, the turbo receiver achieves the best BLER performance. Assuming the block error rate requirement is $10^{-3}$, the turbo receiver is able to support 40 active users while the separate design can only support 20, which is a remarkable 100\% increase. Compared with the baseline schemes, it also greatly narrows the performance gap to the upper bound scheme with perfect knowledge of the user activity, owing to the more accurate activity detection and channel estimation. Because of the same reason, the SI-aided receiver brings notable BLER reduction compared with the data-assisted design.

\begin{figure}[htbp]
\centering
\includegraphics[width=3in]{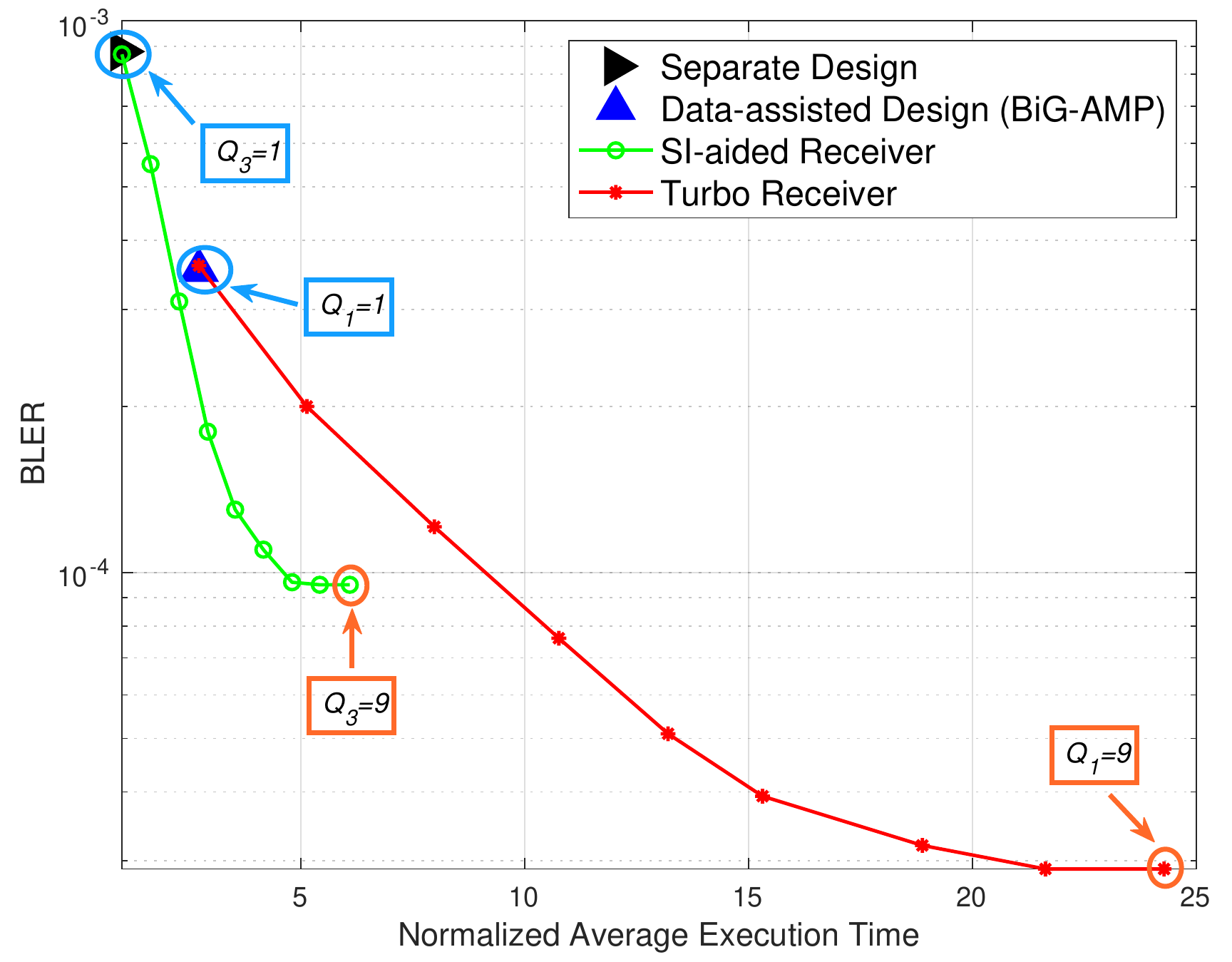}
\caption{BLER vs. the normalized average execution time ($K=20$).}
\label{S4}
\end{figure}

Since both the turbo receiver and the SI-aided receiver iterate between an estimator and a channel decoder, we further investigate the impacts of the number of iterations, i.e., $Q_{1}$ for the turbo receiver and $Q_{3}$ for the SI-aided receiver, as shown in Fig. \ref{S4}. We examine the computation complexity of different schemes by measuring their average execution time on the same computing server. Since the average execution time is platform-specific, it is normalized with respect to that of the separate design. As shown in the figure, the separate design has the lowest complexity but the highest BLER, as it ignores both the common sparsity pattern and the information offered by the channel decoder. Besides, it is observed that the performance achieved by both of the proposed receivers improves with the number of iterations (i.e., $Q_{1}$ for the turbo receiver and $Q_{3}$ for the data-assisted design), which again corroborates the effectiveness of the iterative estimation on the prior information of the user activity and the transmitted data symbols. However, such performance improvement is accompanied with increased computation complexity. Compared with the turbo receiver, the SI-aided receiver enjoys 66\%$\sim$74\% average execution time reduction since the sequential estimator only processes the pilot signal for JADCE, and channel decoding is performed just for the users that have not passed the parity check. In addition, we notice that the major performance gains of the proposed receivers come from the first few iterations, e.g., seven in the considered scenario, and the subsequent iterations only contribute to marginal further improvement. In other words, there is no need to execute a large number of iterations, and wise choices of hyper-parameters for the proposed receivers are critical to balance the performance gain and the computation cost.

\begin{figure}[htbp]
\centering
\includegraphics[width=3in]{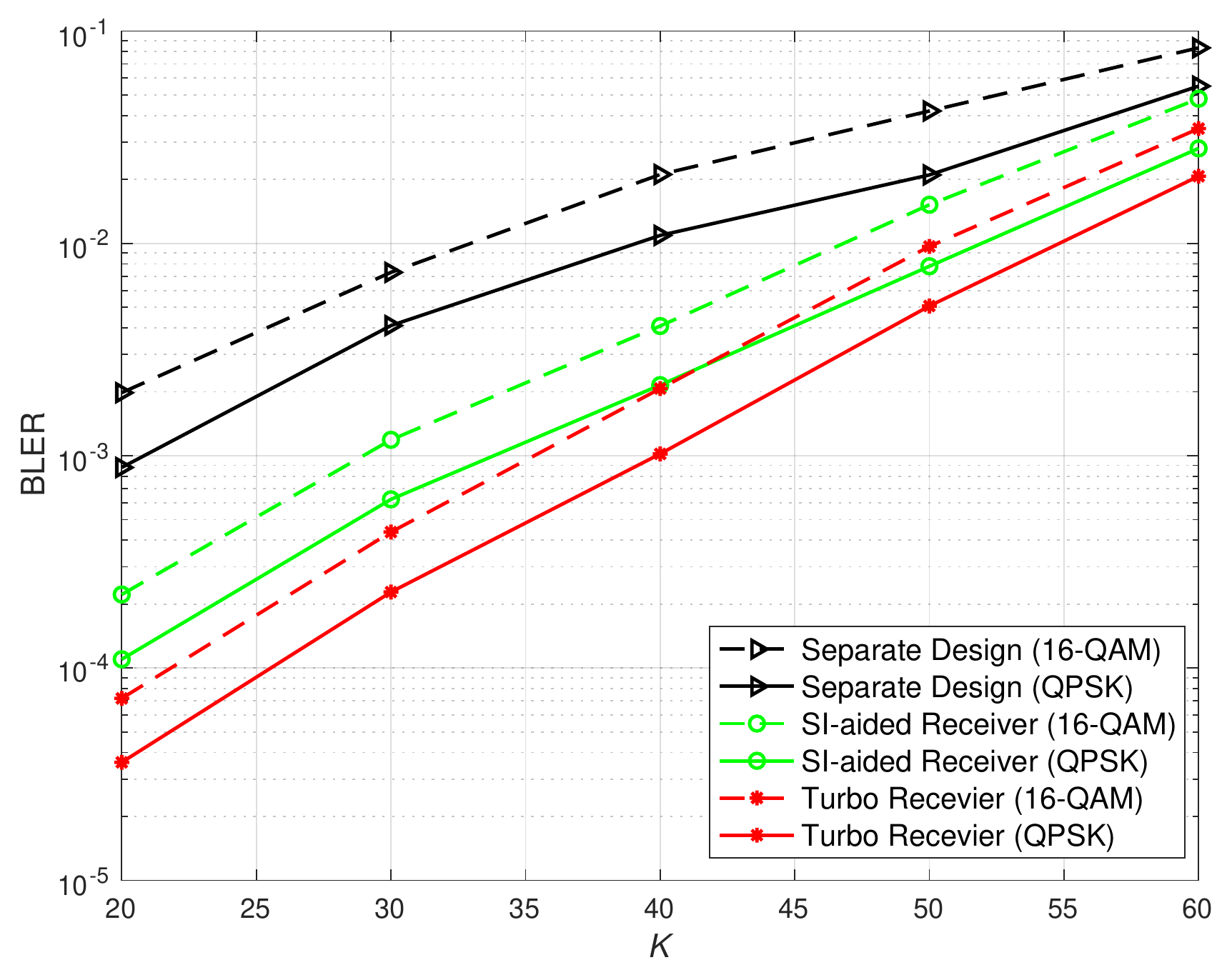}
\caption{BLER vs. the number of active users with QPSK and 16-QAM.}
\label{Modulation}
\end{figure}

To show the effectiveness of the two proposed receivers with different modulation schemes, we also simulate a grant-free massive RA system with 16-QAM while keeping $L_{d}$ unchanged. As shown in Fig. \ref{Modulation}, when 16-QAM is used, the BLER performance of different receivers degrades, which is in accordance with intuition. However, we see the two proposed receivers still achieve significant performance improvements compared with the separate design, which again demonstrates the benefit of utilizing the common sparsity and data decoding results in designing grant-free massive RA receivers.

\begin{figure}[htbp]
\centering
\includegraphics[width=3in]{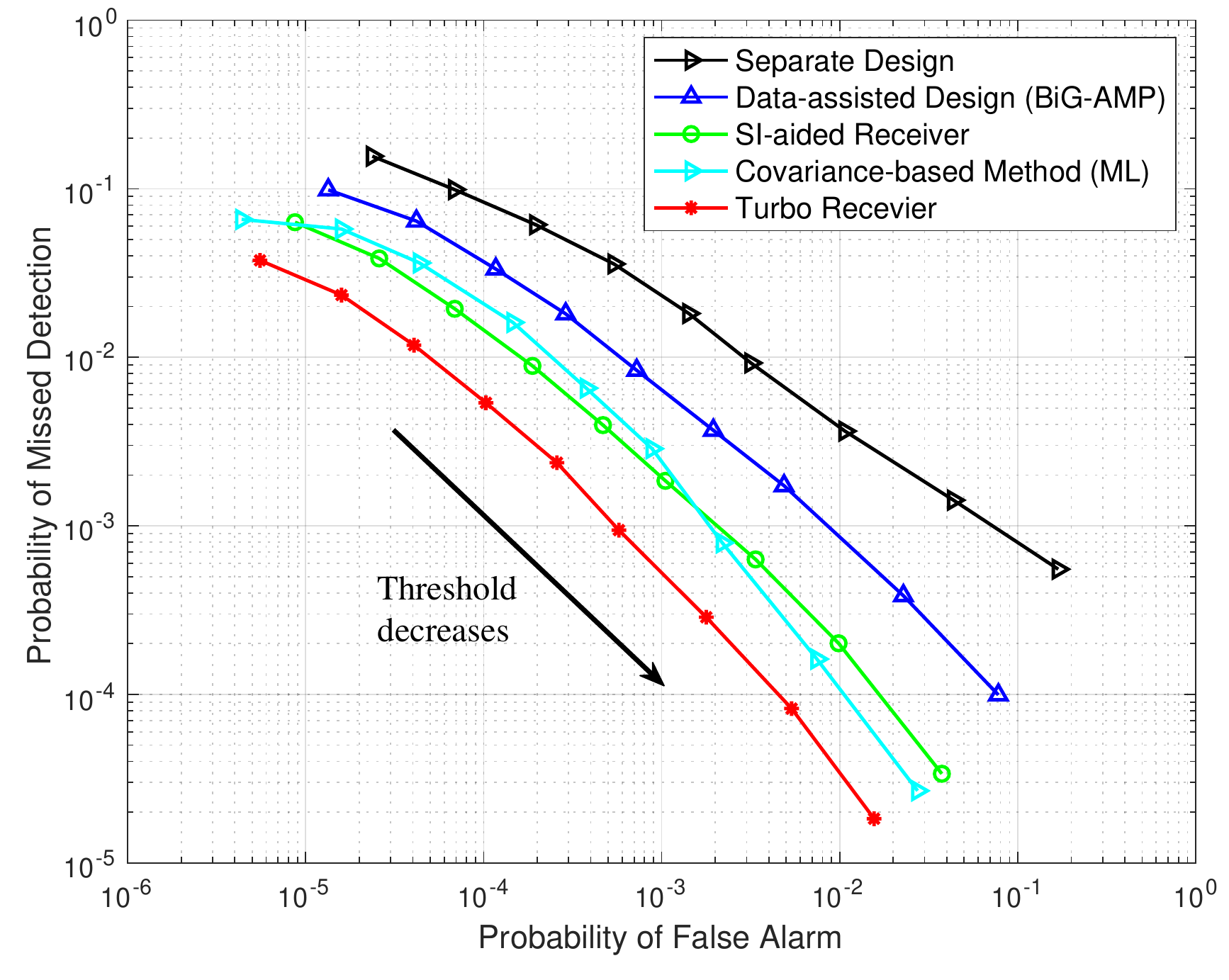}
\caption{Probability of missed detection vs. probability of false alarm ($K=30$, $\theta \in \{0.1, 0.2, 0.3, 0.4, 0.5, 0.6, 0.7, 0.8, 0.9\}$, and $\nu \in \{3, 10, 20, 35, 50, 75, 100, 200, 300\}$).}
\label{ROC}
\end{figure}

\begin{figure}[htbp]
\centering
\includegraphics[width=3in]{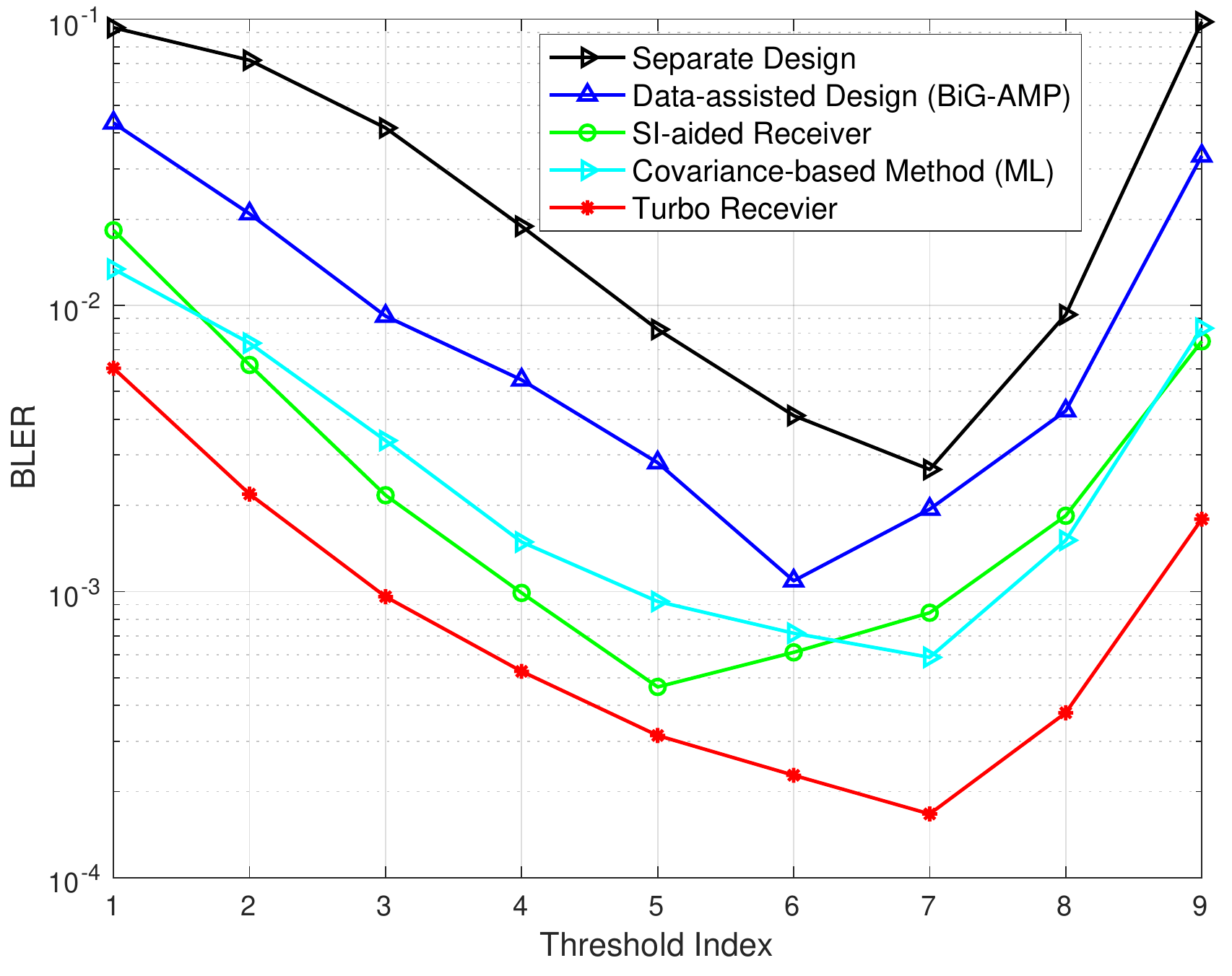}
\caption{BLER vs. threshold index ($K=30$, $\theta \in \{0.1, 0.2, 0.3$, $0.4, 0.5, 0.6, 0.7, 0.8, 0.9\}$, and $\nu \in \{3, 10, 20, 35, 50, 75, 100, 200, 300\}$.}
\label{ROCBLER}
\end{figure}

Furthermore, we investigate the impacts of the threshold value $\theta$ for determining the set of active users in the AMP-based receivers. A covariance-based receiver is also simulated, which applies the covariance-based method \cite{shag2018} for activity detection together with an MMSE channel estimator. Similar to the AMP-based receivers, a threshold $\nu$ ($\nu>0$) was introduced to detect the set of active users in \cite{shag2018}. It is observed from Fig. \ref{ROC} and Fig. \ref{ROCBLER} that there is a tradeoff between the missed detection and false alarm probability, and an optimal threshold value brings the best BLER performance for a given receiver. The two proposed receivers achieve better performance compared with the baselines with various values of $\theta$. Although the covariance-based receiver outperforms the conventional separate and data-assisted designs, it is defeated by the proposed turbo receiver by a large margin. On the other hand, while the SI-aided receiver achieves comparable activity detection performance as the covariance-based receiver, it outperforms the covariance-based receiver in terms of BLER with an optimized threshold value. It is also worthwhile to note that the covariance-based receiver has a very high complexity and its execution time is 6.5 times of the separate design, but that of the SI-aided receiver is only $4.2$ times. These results demonstrate the competence of the proposed receivers over the covariance-based receiver.

\section{Conclusions}
This paper carried out the first holistic investigation that jointly considered activity detection, channel estimation, and data decoding for grant-free massive random access (RA). A turbo receiver was proposed to exploit the common sparsity pattern in the received pilot and data signal, and its performance is enhanced by the extrinsic information from the channel decoder. To reduce the complexity, we also developed a low-cost side information (SI)-aided receiver, where the SI is updated iteratively to take advantages of the common sparsity pattern and the channel decoding results. Simulation results demonstrated that substantial performance gains can be obtained with advanced receivers for a given protocol.

Our study demonstrates the benefits of exploiting the structured information from the pilot and data symbols, and the importance of incorporating the interplay among activity detection, channel estimation, and data decoding to the design of massive RA receivers. In other words, treating activity detection and data decoding separately, as in many previous studies, leads to highly suboptimal receivers. For future investigations, it would be interesting to extend the proposed receivers to scenarios with spatial-temporal correlation of the user activity and investigate more complex massive RA systems supported by ultra-massive MIMO and reconfigurable meta-surfaces. Meanwhile, since the proposed receivers are iterative algorithms by nature, it is critical to further reduce their computational complexity via, for examples, deep learning based methods \cite{vgsa2021,yshen2021} to facilitate practical implementation. In addition, optimally controlling the uplink transmit power is also very important.

\appendices
\section{Derivations of $\hat{z}_{mt}$ and Its Variance}
Since $p(y_{mt}|z_{mt}) = \mathcal{CN}\left(z_{mt};y_{mt},\frac{\sigma^{2}}{\gamma}\right)$ and the prior distribution of $p(z_{mt})$ is approximated as $\mathcal{CN}\Big(z_{mt};M_{mt}^{p(j)}(i),$ $V_{mt}^{p(j)}(i)\Big)$, the joint PDF $p(y_{mt},z_{mt})$ is approximated in the $i$-th iteration of Algorithm 2 by
\begin{align}
J_{z_{mt}}^{(j)}(i)&= \mathcal{CN}\left(z_{mt};y_{mt},\frac{\sigma^{2}}{\gamma}\right)\mathcal{CN}\left(z_{mt};M_{mt}^{p(j)}(i),V_{mt}^{p(j)}(i)\right)\\\notag
&= A_{1}\cdot \mathcal{CN}\left(z_{mt};C^{(j)}(i),D^{(j)}(i)\right),
\end{align}
\noindent where $C^{(j)}(i) \triangleq \frac{y_{mt}V_{mt}^{p(j)}(i)\gamma+\sigma^{2}M_{mt}^{p(j)}(i)}{\sigma^{2}+V_{mt}^{p(j)}(i)\gamma}$, $D^{(j)}(i)\triangleq \frac{\sigma^{2}V_{mt}^{p(j)}(i)}{\sigma^{2}+V_{mt}^{p(j)}(i)\gamma}$, and $A_{1} \triangleq \mathcal{CN}\Big(0;y_{mt}-M_{mt}^{p(j)}(i),\frac{\sigma^{2}}{\gamma}+V_{mt}^{p(j)}(i)\Big)$. Thus, the posterior distribution of $z_{mt}$ can be approximated in the $i$-th iteration via the Bayes' rule as follows:
\begin{align}
r_{z_{mt}}^{(j)}(i) = \frac{J_{z_{mt}}^{(j)}(i)}{\int J_{z_{mt}}^{(j)}(i) dz_{mt}}
=\mathcal{CN}\left(z_{mt};C^{(j)}(i),D^{(j)}(i)\right).
\end{align}

\noindent Since the MMSE estimate of $z_{mt}$ is the posterior mean, we have $\hat{z}_{mt}^{(j)}(i) =\frac{y_{mt}V_{mt}^{p(j)}(i)\gamma+\sigma^{2}M_{mt}^{p(j)}(i)}{\sigma^{2}+V_{mt}^{p(j)}(i)\gamma}$. Accordingly, the variance of the MMSE estimate is given by the posterior variance as $V_{mt}^{z(j)}(i) =\frac{\sigma^{2}V_{mt}^{p(j)}(i)}{\sigma^{2}+V_{mt}^{p(j)}(i)\gamma}$.

\section{Derivations of $P_{mn}^{h\left(j\right)}\left(i\right)$ and $Q_{mn}^{h\left(j\right)}\left(i\right)$}
According to the principles of the BiG-AMP algorithm, $\prod_{t=1}^{L} I_{f_{y_{mt}} \rightarrow h_{mn}}^{(j)}\left(i\right)$ and $\prod_{t=L+1}^{T} I_{f_{y_{mt}} \rightarrow h_{mn}}^{(j)}\left(i\right)$ are approximated as complex Gaussian distributions $\mathcal{CN}\Big(h_{mn};P_{p,mn}^{h(j)}(i),$ $Q_{p,mn}^{h(j)}(i)\Big)$ and $\mathcal{CN}\Big(h_{mn};P_{d,mn}^{h(j)}(i), Q_{d,mn}^{h(j)}(i)\Big)$, respectively. By substituting these two complex Gaussian PDFs into the term $\prod_{t=1}^{L} I_{f_{y_{mt}} \rightarrow h_{mn}}^{(j)}\left(i\right)\prod_{t=L+1}^{T} I_{f_{y_{mt}} \rightarrow h_{mn}}^{(j)}\left(i\right)$, we have:
\begin{align}
    &\prod_{t=1}^{L} I_{f_{y_{mt}} \rightarrow h_{mn}}^{(j)}\left(i\right)\prod_{t=L+1}^{T} I_{f_{y_{mt}} \rightarrow h_{mn}}^{(j)}\left(i\right)\\\notag &\!=\!\mathcal{CN}\!\left(h_{mn};\!P_{p,mn}^{h(j)}(i),Q_{p,mn}^{h(j)}(i)\!\right)\!\mathcal{CN}\!\left(h_{mn};\!P_{d,mn}^{h(j)}(i),Q_{d,mn}^{h(j)}(i)\!\right)\\\notag
    &\!=\!A_{2} \cdot \mathcal{CN}\!\left(h_{mn};P_{mn}^{h(j)}(i),Q_{mn}^{h(j)}(i)\right),
\end{align}
    
\noindent where $A_{2}\!\triangleq\! \mathcal{CN}\left(0;P_{p,mn}^{h(j)}(i)\!-\!P_{d,mn}^{h(j)}(i),P_{p,mn}^{h(j)}(i)\!+\!P_{d,mn}^{h(j)}(i)\right)$, $P_{m n}^{h(j)}(i)=\frac{P_{p,m n}^{h(j)}(i)Q_{d,m n}^{h(j)}(i)+P_{d,m n}^{h(j)}(i)Q_{p,m n}^{h(j)}(i)}{Q_{p,m n}^{h(j)}(i)+Q_{d,m n}^{h(j)}(i)}$, and $Q_{m n}^{h(j)}(i)=$ $\frac{Q_{p,m n}^{h(j)}(i)Q_{d,m n}^{h(j)}(i)}{Q_{p,m n}^{h(j)}(i)+Q_{d,m n}^{h(j)}(i)}$.

\section{Derivation of (16)}
It is straightforward that $p\left(h_{mn}|u_{n}=1\right) = \mathcal{CN}(h_{mn};$ $0,\beta_{n})$ and $p\left(h_{mn}|u_{n}=0\right) = \delta(h_{mn})$. Thus, based on the BP algorithm, the term $I_{f_{h_{mn} \rightarrow h_{mn}}}^{(j)}(i)$ can be determined as
\begin{align}
I_{f_{h_{mn}}\rightarrow h_{mn}}^{(j)}(i)&=\sum_{u_{n} \in \{0,1\}}p(h_{mn}|u_{n})I_{u_{n}\rightarrow f_{h_{mn}}}^{(j)}(i)\\\notag
&= I_{u_{n}\rightarrow f_{h_{mn}}}^{\left(j\right)}\left(i\right) \Big |_{u_{n}=0}\delta(h_{mn})\\\notag
&+I_{u_{n}\rightarrow f_{h_{mn}}}^{\left(j\right)}\left(i\right) \Big |_{u_{n}=1}\mathcal{CN}(h_{mn};0,\beta_{n}),
\end{align}

\noindent where $I_{u_{n} \rightarrow f_{h_{mn}}}^{(j)}(i)$ is the message from variable node $u_{n}$ to factor node $p(h_{mn}|u_{n})$ that can be obtained as follows:
\begin{align}
    I_{u_{n} \rightarrow f_{h_{mn}}}^{(j)}(i)=p(u_{n}) \prod_{k \in \mathcal{M} \backslash \{m\}} I_{f_{h_{kn}} \rightarrow u_{n}}^{(j)}(i).
\end{align}

\noindent In (34), $p\left(u_{n}\right)$ stands for the likelihood that user $n$ is active or not in the considered transmission block, and $I_{f_{h_{kn}} \rightarrow u_{n}}^{(j)}(i)$ is the message from factor node $p(h_{kn}|u_{n})$ ($k \in\mathcal{M}\setminus \{m\}$) to variable node $u_{n}$ that can be expanded as follows:
\begin{align}
    I_{f_{h_{kn}}\rightarrow u_{n}}^{(j)}(i) = \int p\left(h_{kn}|u_{n}\right) I_{h_{kn} \rightarrow f_{h_{kn}}}^{(j)}(i)dh_{kn}.
\end{align}

\noindent In (35), $I_{h_{kn} \rightarrow f_{h_{kn}}}^{(j)}(i)$ is the message from variable node $h_{kn}$ to factor node $p(h_{kn}|u_{n})$ that can be obtained based on the BP algorithm as follows:
\begin{align}
    I_{h_{kn} \rightarrow f_{h_{kn}}}^{(j)}(i)=\prod_{t=1}^{L} I_{f_{y_{kt}} \rightarrow h_{kn}}^{(j)}\left(i\right)\prod_{t=L+1}^{T} I_{f_{y_{kt}} \rightarrow h_{kn}}^{(j)}\left(i\right)\\\notag
    = A_{3} \cdot \mathcal{CN}\left(h_{kn};P_{kn}^{h(j)}(i),Q_{kn}^{h(j)}(i)\right),
\end{align}

\noindent where the second equality in (36) adopts the same approximation as the one used in (32), and $A_{3}$ is a constant given as $\mathcal{CN}\Big(0;$ $P_{p,kn}^{h(j)}(i)-P_{d,kn}^{h(j)}(i),Q_{p,kn}^{h(j)}(i)+Q_{d,kn}^{h(j)}(i)\Big)$. By substituting the right-hand side of (36) into (35), we have:
\begin{align}
    I_{f_{h_{kn}}\rightarrow u_{n}}^{(j)}(i) &= A_{3} \int p(h_{kn}|u_{n})\\\notag
    &\times \mathcal{CN} \left(h_{kn};P_{kn}^{h(j)}(i),Q_{kn}^{h(j)}(i)\right)dh_{kn}\\\notag
&=A_{3} \left\{
\begin{array}{ll}
\mathcal{CN}\left(0;P_{kn}^{h(j)}(i), Q_{kn}^{h(j)}(i)+\beta_{n}\right),  u_{n}=1, \\
\mathcal{CN}\left(0;P_{kn}^{h(j)}(i), Q_{kn}^{h(j)}(i)\right),  u_{n}=0.
\end{array}\right.
\end{align}
\noindent Define
\begin{align}
L_{mn}^{(j)}(i) \triangleq \ln\left(\frac{I_{u_{n} \rightarrow f_{h_{mn}}}^{(j)}(i)|_{u_{n}=1}}{I_{u_{n} \rightarrow f_{h_{mn}}}^{(j)}(i)|_{u_{n}=0}}\right),
\end{align}

\noindent which uniquely determines the values of $I_{u_{n} \rightarrow f_{h_{mn}}}^{(j)}(i) \Big|_{u_{n}=1}$ and $I_{u_{n} \rightarrow f_{h_{mn}}}^{(j)}(i) \Big|_{u_{n}=0}$, since $\sum_{u_{n}\in\{0,1\}} I_{u_{n} \rightarrow f_{h_{mn}}}^{(j)}(i) = 1$. We further define
\begin{align}
K_{kn}^{(j)}(i)&\triangleq \ln\left(\frac{I_{f_{h_{kn}}\rightarrow u_{n}}^{(j)}(i)|_{u_{n}=1}}{I_{f_{h_{kn}}\rightarrow u_{n}}^{(j)}(i)|_{u_{n}=0}}\right)\\\notag
&= \ln\left(\frac{Q_{m n}^{h(j)}(i)}{Q_{m n}^{h(j)}(i)+\beta_{n}}\right)+\frac{|P_{mn}^{h(j)}(i)|^{2}\beta_{n}}{\left(Q_{m n}^{h(j)}(i)+\beta_{n}\right)Q_{m n}^{h(j)}(i)}
\end{align}
\noindent and $U_{n}\triangleq \ln\left(\frac{p\left(u_{n}=1\right)}{p\left(u_{n}=0\right)}\right)$. With some basic mathematical manipulations, $L_{mn}^{(j)}(i)$ can be simplified as follows:
\begin{align}
L_{mn}^{(j)}(i)= U_{n} + \sum\nolimits_{k \in \mathcal{M}\backslash \{m\}} K_{kn}^{(j)}(i).
\end{align}

\noindent As a result, the terms $I_{u_{n}\rightarrow f_{h_{mn}}}^{(j)}(i) \Big|_{u_{n}=1}$ and $I_{u_{n}\rightarrow f_{h_{mn}}}^{(j)}(i) \Big|_{u_{n}=0}$ can be determined as $\frac{\exp{\left(L_{mn}^{(j)}(i)\right)}}{1+\exp{\left(L_{mn}^{(j)}(i)\right)}}$ and $\frac{1}{1+\exp{\left(L_{mn}^{(j)}(i)\right)}}$, respectively. We complete the derivation by defining $\rho_{mn}^{(j)}(i)\triangleq \frac{\exp{\left(L_{mn}^{(j)}(i)\right)}}{1+\exp{\left(L_{mn}^{(j)}(i)\right)}}$.
\ifCLASSOPTIONcaptionsoff
  \newpage
\fi


\begin{thebibliography}{1}
\bibitem{xbian20211} X. Bian, Y. Mao, and J. Zhang, “Joint activity detection and data decoding in massive random access via a turbo receiver,” in \emph{Proc. IEEE Int. Workshop Signal Process. Adv. Wireless Commun. (SPAWC)}, Sep. 2021.
\bibitem{aal2015} A. Al-Fuqaha, M. Guizani, M. Mohammadi, M. Aledhari, and M. Ayyash, “Internet of Things: A survey on enabling technologies, protocols, and applications,” \emph{IEEE Commun. Surveys Tut.}, vol. 17, no. 4, pp. 2347-2376, Fourth Quart. 2015.
\bibitem{Cisco} Cisco, “Cisco annual Internet report (2018–2023),” \emph{Cisco White Paper}, Mar. 2020.
\bibitem{itu} ITU-R, “Framework and overall objectives of the future development of IMT for 2020 and beyond Recommendation,” \emph{ITU-R M.2083-0}, 2015.
\bibitem{cboc2016} C. Bockelmann, N. Pratas, H. Nikopour, K. Au, T. Svensson, C. Stefanovic, P. Popovski, and A. Dekorsy, “Massive machine-type communications in 5G: Physical and MAC-layer solutions,” \emph{IEEE Commun. Mag.}, vol. 54, no. 9, pp. 59–65, Sep. 2016.
\bibitem{mha2013} M. Hasan, E. Hossain, and D. Niyato, “Random access for machine-to-machine communication in LTE-advanced networks: Issues and approaches,” \emph{IEEE Commun. Mag.}, vol. 51, no. 6, pp. 86-93, Jun. 2013.
\bibitem{ebj2017} E. Björnson, E. Carvalho, J. H. Sørensen, E. G. Larsson, and P. Popovski, “A random access protocol for pilot allocation in crowded massive MIMO systems,” \emph{IEEE Trans. Wireless Commun.}, vol. 16, no. 4, pp. 2220-2234, Apr. 2017.
\bibitem{njiang2018} N. Jiang, Y. Deng, A. Nallanathan, X. Kang, and T. Q. S. Quek, “Analyzing random access collisions in massive IoT networks,” \emph{IEEE Trans. Wireless Commun.}, vol. 17, no. 10, pp. 6853-6879, Oct. 2018.
\bibitem{lliu2018} L. Liu, E. G. Larsson, W. Yu, P. Popovski, C. Stefanović, and E. Carvalh, “Sparse signal processing for grant-free massive connectivity: A future paradigm for random access protocols in the Internet of Things,” \emph{IEEE Signal Process. Mag.}, vol. 35, no. 5, pp. 88-99, Sep. 2018.
\bibitem{psh2017} P. Schulz \emph{et al.}, “Latency critical IoT applications in 5G: Perspective on the design of radio interface and network architecture,” \emph{IEEE Commun. Mag.}, vol. 55, no. 2, pp. 70-78, Feb. 2017.
\bibitem{xchen2021} X. Chen, D. Ng, W. Yu, E. G. Larsson, N. Al-Dhahir, and R. Schober, “Massive access for 5G and beyond,” \emph{IEEE J. Sel. Areas Commun.}, vol. 39, no. 3, pp. 615-637, Mar. 2021.
\bibitem{ywu2020} Y. Wu, X. Gao, S. Zhou, W. Yang, Y. Polyanskiy, and G. Caire, “Massive access for future wireless communication systems,” \emph{IEEE Wireless Commun.}, vol. 27, no. 4, pp. 148-156, Aug. 2020.
\bibitem{donoho2006} D. L. Donoho, “Compressed sensing,” \emph{IEEE Trans. Inf. Theory}, vol. 52, no. 4, pp. 1289–1306, Apr. 2006.
\bibitem{trob1996} T. Robert, “Regression shrinkage and selection via the Lasso,” \emph{J. Roy. Statist. Soc.}, vol. 58, no. 1, pp. 267–288, Jan. 1996.
\bibitem{jat2004} J. Tropp, “Greed is good: Algorithmic results for sparse approximation,” \emph{IEEE Trans. Inf. Theory}, vol. 50, no. 10, pp. 2231– 2242, Oct. 2004.
\bibitem{donoho2010} D. L. Donoho, A. Maleki, and A. Montanari, “Message passing algorithms for compressed sensing: I. motivation and construction,” in \emph{Proc. IEEE Inf. Theory Workshop (ITW)}, Cairo, Egypt, Jan. 2010.
\bibitem{jtpa2014} J. T. Parker, P. Schniter, and V. Cevher, “Bilinear generalized approximate message passing – Part I: Derivation,” \emph{IEEE Trans. Signal Process.}, vol. 62, no. 22, pp. 5839–5853, Nov. 2014.
\bibitem{shag2018} S. Haghighatshoar, P. Jung, and G. Caire, “Improved scaling law for activity detection in massive MIMO systems,” in \emph{Proc. IEEE Int. Symp. Inf. Theory (ISIT)}, Vail, CO, USA, Jun. 2018.
\bibitem{zchen2019} Z. Chen, F. Sohrabi, and W. Yu, “Multi-cell sparse activity detection for massive random access: Massive MIMO versus cooperative MIMO,” \emph{IEEE Trans. Wireless Commun.}, vol. 18, no. 8, pp. 4060-4072, Aug. 2019.
\bibitem{bwang2016} B. Wang, L. Dai, Y. Zhang, T. Mir, and J. Li, “Dynamic compressive sensing-based multi-user detection for uplink grant-Free NOMA,” \emph{IEEE Commun. Lett.}, vol. 20, no. 11, pp. 2320-2323, Nov. 2016.
\bibitem{cwei2017} C. Wei, H. Liu, Z. Zhang, J. Dang, and L. Wu, “Approximate message passing-based joint user activity and data detection for NOMA,” \emph{IEEE Commun. Lett.}, vol. 21, no. 3, pp. 640–643, Mar. 2017.
\bibitem{zchen2018} Z. Chen, F. Sohrabi, and W. Yu, “Sparse activity detection for massive connectivity,” \emph{IEEE Trans. Signal Process.}, vol. 66, no. 7, pp. 1890–1904, Apr. 2018.
\bibitem{lliuwyu2018} L. Liu and W. Yu, “Massive connectivity with massive MIMO – Part I: Device activity detection and channel estimation,” \emph{IEEE Trans. Signal Process.}, vol. 66, no. 11, pp. 2933–2946, Jun. 2018.
\bibitem{mke2020} M. Ke, Z. Gao, Y. Wu, X. Gao, and R. Schober, “Compressive sensing based adaptive active user detection and channel estimation: Massive access meets massive MIMO,” \emph{IEEE Trans. Signal Process.}, vol. 68, pp. 764–779, 2020.
\bibitem{ycheng2021} Y. Cheng, L. Liu, and P. Li, “Orthogonal AMP for massive access in channels with spatial and temporal correlations,” \emph{IEEE J. Sel. Areas Commun.}, vol. 39, no. 3, pp. 726-740, Mar. 2021.
\bibitem{xshao2019} X. Shao, X. Chen and R. Jia, “A dimension reduction-based joint activity detection and channel estimation algorithm for massive access,” \emph{IEEE Trans. Signal Process.}, vol. 68, pp. 420-435, Dec. 2019.
\bibitem{ycui2021} Y. Cui, S. Li and W. Zhang, “Jointly sparse signal recovery and support recovery via deep learning with applications in MIMO-based grant-free random access,” \emph{IEEE J. Sel. Areas Commun.}, vol. 39, no. 3, pp. 788-803, Mar. 2021.
\bibitem{ydu2018} Y. Du, \emph{et al.}, “Joint channel estimation and multiuser detection for uplink grant-free NOMA,” \emph{IEEE Wireless Commun. Lett.}, vol. 7, no. 4, pp. 682–685, Feb. 2018.
\bibitem{qzou2020} Q. Zou, H. Zhang, D. Cai and H. Yang, “A low-complexity joint user activity, channel and data estimation for grant-free massive MIMO systems,” \emph{IEEE Signal Process. Lett.}, vol. 27, pp. 1290-1294, 2020.
\bibitem{xbian2021} X. Bian, Y. Mao, and J. Zhang, “Supporting more active users for massive access via data-assisted activity detection,” in \emph{Proc. IEEE Int. Conf. Commun. (ICC)}, Montreal, QC, Canada, Jun. 2021.
\bibitem{bmho2003} B. M. Hochwald and S. t. Brink, “Achieving near-capacity on a multiple-antenna channel,” \emph{IEEE Trans. Commun.}, vol. 51, no. 3, pp. 389-399, Mar. 2003.
\bibitem{kkwong2007} K. Wong, A. Paulraj, and R. Murch, “Efficient high-performance decoding for overloaded MIMO antenna systems,” \emph{IEEE Trans. Wireless Commun.}, vol. 6, no. 5, pp. 1833–1843, May 2007.
\bibitem{3gpp2011} 3GPP, “3GPP TS 36.212 version 10.0.0 Release 10,” Jan. 2011.
\bibitem{3gpp2020} 3GPP, “3GPP TS 38.212 version 15.10.0 Release 15,” Nov. 2020.
\bibitem{tcui2006} T. Cui and C. Tellambura, “Power delay profile and noise variance estimation for OFDM,” \emph{IEEE Commun. Lett.}, vol. 10, no. 1, pp. 25-27, Jan. 2006.
\bibitem{shay2004} S. Haykin, M. Sellathurai, Y. de Jong, and T. Willink, “Turbo-MIMO for wireless communications,” \emph{IEEE Commun. Mag.}, vol. 42, no. 10, pp. 48-53, Oct. 2004.
\bibitem{xwa2004} X. Wautelet, A. Dejonghe, and L. Vandendorpe, “MMSE-based fractional turbo receiver for space-time BICM over frequency-selective MIMO fading channels,” \emph{IEEE Trans. Signal Process.}, vol. 52, no. 6, pp. 1804-1809, Jun. 2004.
\bibitem{cbe1996} C. Berrou and A. Glavieux, “Near optimum error correcting coding and decoding: Turbo-codes,” \emph{IEEE Trans. Commun.}, vol. 44, no. 10, pp. 1261-1271, Oct. 1996.
\bibitem{frks2001} F. R. Kschischang, B. J. Frey and H. A. Loeliger, “Factor graphs and the sum-product algorithm,” \emph{IEEE Trans. Inf. Theory}, vol. 47, no. 2, pp. 498-519, Feb. 2001.
\bibitem{smkay1993} S. Kay,\! \emph{Fundamentals of Statistical Signal Processing:\! Estimation Theory}. Prentice Hall, Englewood Cliffs, NJ, USA, 1993.
\bibitem{rssu2018} R. S. Sutton and A. G. Barto, \emph{Reinforcement Learning: An Introduction}. MIT Press, 2018.
\bibitem{bvu2001} B. Vucetic and J. Yuan, \emph{Turbo Codes: Principles and Applications}. Springer, 2001.
\bibitem{rgga1962} R. G. Gallager, “Low density parity check codes,” \emph{IRE Trans. Inf. Theory}, vol. IT-8, no. 1, pp. 21-28, Jan. 1962.
\bibitem{hoc2004} D. E. Hocevar, “A reduced complexity decoder architecture via layered decoding of LDPC codes,” in \emph{Proc. IEEE Workshop Signal Process. Syst. (SiPS)}, Austin, TX, USA, Oct. 2004.
\bibitem{jha1996} J. Hagenauer, E. Offer, and L. Papke, “Iterative decoding of binary block and convolutional codes,” \emph{IEEE Trans. Inf. Theory}, vol. 42, no. 2, pp. 429-445, Mar. 1996.
\bibitem{ama2019} A. Ma, Y. Zhou, C. Rush, D. Baron, and D. Needell, “An approximate message passing framework for side information,” \emph{IEEE Trans. Signal Process.}, vol. 67, no. 7, pp. 1875-1888, Apr. 2019. 
\bibitem{bgo2018} B. Goektepe, S. Faehse, L. Thiele, T. Schierl, and C. Hellge, “Subcode-based early HARQ for 5G,” in \emph{Proc. IEEE Int. Conf. Commun. (ICC)}, Kansas City, MO, USA, May 2018.
\bibitem{3gpp2018} 3GPP, “3GPP TS 36.211 version 15.3.0 Release 15,” Oct. 2018.
\bibitem{vgsa2021} V. Satorras and M. Welling, “Neural enhanced belief propagation on factor graphs,” in \emph{Proc. AISTATS-21}, pp. 685–693, Apr. 2021.
\bibitem{yshen2021} Y. Shen, Y. Shi, J. Zhang, and K. B. Letaief, “Graph neural networks for scalable radio resource management: architecture design and theoretical analysis,” \emph{IEEE J. Select. Areas Commun.}, vol. 39, no. 1, pp. 101–115, Jan. 2021.
\end{thebibliography}
\end{document}